# Hydrogen peroxide electrogeneration from $O_2$ electroreduction: a review focusing on carbon electrocatalysts and environmental applications

Aline B. Trench[1], Caio Machado Fernandes[1], João Paulo C. Moura[1], Lanna E. B. Lucchetti[1], Thays S. Lima[2], Vanessa S. Antonin[1], James M. de Almeida[1], Pedro Autreto[1], Irma Robles[3], Artur J. Motheo[2], Marcos R. V. Lanza[2], Mauro C. Santos[1*]

[1] *Centre of Natural and Human Sciences, Federal University of ABC. Rua Santa Adélia 166, Bairro Bangu, 09210-170, Santo André - SP, Brazil.*

[2] *São Carlos Institute of Chemistry, University of São Paulo, P.O. Box 780, São Carlos, SP CEP 13560-970, Brazil.*

[3] *Center for Research and Technological Development in Electrochemistry, S.C., Parque Tecnologico Queretaro, 76703, Sanfandila, Pedro Escobedo, Queretaro, Mexico.*

*Corresponding Author:*

**E-mail:* mauro.santos@ufabc.edu.br

# Abstract

Hydrogen peroxide ($H_2O_2$) stands as one of the foremost utilized oxidizing agents in modern times. The established method for its production involves the intricate and costly anthraquinone process. However, a promising alternative pathway is the electrochemical hydrogen peroxide production, accomplished through the oxygen reduction reaction via a 2-electron pathway. This method not only simplifies the production process but also upholds environmental sustainability, especially when compared to the conventional anthraquinone method. In this review paper, recent works from the literature focusing on the 2-electron oxygen reduction reaction promoted by carbon electrocatalysts are summarized. The practical applications of these materials in the treatment of effluents contaminated with different pollutants (drugs, dyes, pesticides, and herbicides) are presented. Water treatment aiming to address these issues can be achieved through advanced oxidation electrochemical processes such as electro-Fenton, solar-electro-Fenton, and photo-electro-Fenton. These processes are discussed in detail in this work and the possible radicals that degrade the pollutants in each case are highlighted. The review broadens its scope to encompass contemporary computational simulations focused on the 2-electron oxygen reduction reaction, employing different models to describe carbon-based electrocatalysts. Finally, perspectives and future challenges in the area of carbon-based electrocatalysts for $H_2O_2$ electrogeneration are discussed. This review paper presents a forward-oriented viewpoint of present innovations and pragmatic implementations, delineating forthcoming challenges and prospects of this ever-evolving field.

## 1. Introduction

Hydrogen peroxide ($H_2O_2$) is a powerful, versatile, and environmentally friendly oxidizing agent as it decomposes only into water and oxygen (Zhao et al., 2019). It is used for disinfection purposes in hospital environments (Yang et al., 2018), paper and textile bleaching processes (Yu et al., 2017), and chemical syntheses (Puértolas et al., 2015). It can also be employed at water treatment plants based on advanced oxidative processes (AOP), which involve the generation of highly reactive hydroxyl radicals ($^{\bullet}OH$) to break down and remove organic and inorganic pollutants present in wastewater (Cordeiro-Junior et al., 2020). Because it is used in so many different applications, $H_2O_2$ has become one of the 100 most used chemicals in the world (Campos-Martin et al., 2006) and the global demand for this oxidant is expected to grow by 5.7% until 2026 (Pulidindi K).

The main methods of producing $H_2O_2$ are shown in Fig 1. The currently most used industrial method to produce $H_2O_2$ is the anthraquinone oxidation process. This process involves a few steps, namely: (1) reaction of anthraquinone with an alkylating agent to form alkyl anthraquinone derivatives, (2) hydrogenation of the alkyl anthraquinone derivatives to produce hydroquinone derivatives, (3) oxidation of the hydroquinone derivatives to produce hydrogen peroxide as a by-product, and (4) extracting the hydrogen peroxide from the reaction mixture (Campos-Martin et al., 2006; Ingle et al., 2022). The various steps involved in anthraquinone oxidation make this a costly process, requiring adequate infrastructure and significant energy expenditure. Additionally, this process demands safety measures for the handling and storage of large amounts of $H_2O_2$, increasing the cost of $H_2O_2$ production (Li et al., 2022b).

A process that can be used to replace the anthraquinone oxidation method is the direct synthesis of $H_2O_2$ using noble metals such as platinum (Pt) (Edwards et al., 2014) and palladium (Pd) (Wang et al., 2020d) as photocatalysis (Guo et al., 2023). In this case, diluted $H_2O_2$ is generated under UV light irradiation. However, the facilitated $O_2$ evolution is a major disadvantage of these photocatalyst processes, since it compromises the $H_2O_2$ generation. Besides, the high cost of noble metals used in $H_2O_2$ direct synthesis, as well as the risk of explosion caused by the mixture of oxygen and hydrogen, make this method unfeasible for large-scale production (Zhou et al., 2019a; Wang et al., 2021b).

A promising alternative for producing $H_2O_2$ is through the electrochemical method using the oxygen reduction reaction (ORR). This methodology involves the utilization of electricity, water, and air to generate $H_2O_2$ *in situ*, at atmospheric pressure and at moderate temperatures. These conditions set this method apart from the previously mentioned ones due to its economic, safe, and environmentally friendly. As the production is *in situ*, there is no need for storage and transportation, thereby reducing costs and risks associated with $H_2O_2$ production (Stacy et al., 2017; Li et al., 2022b; Wang et al., 2023b).

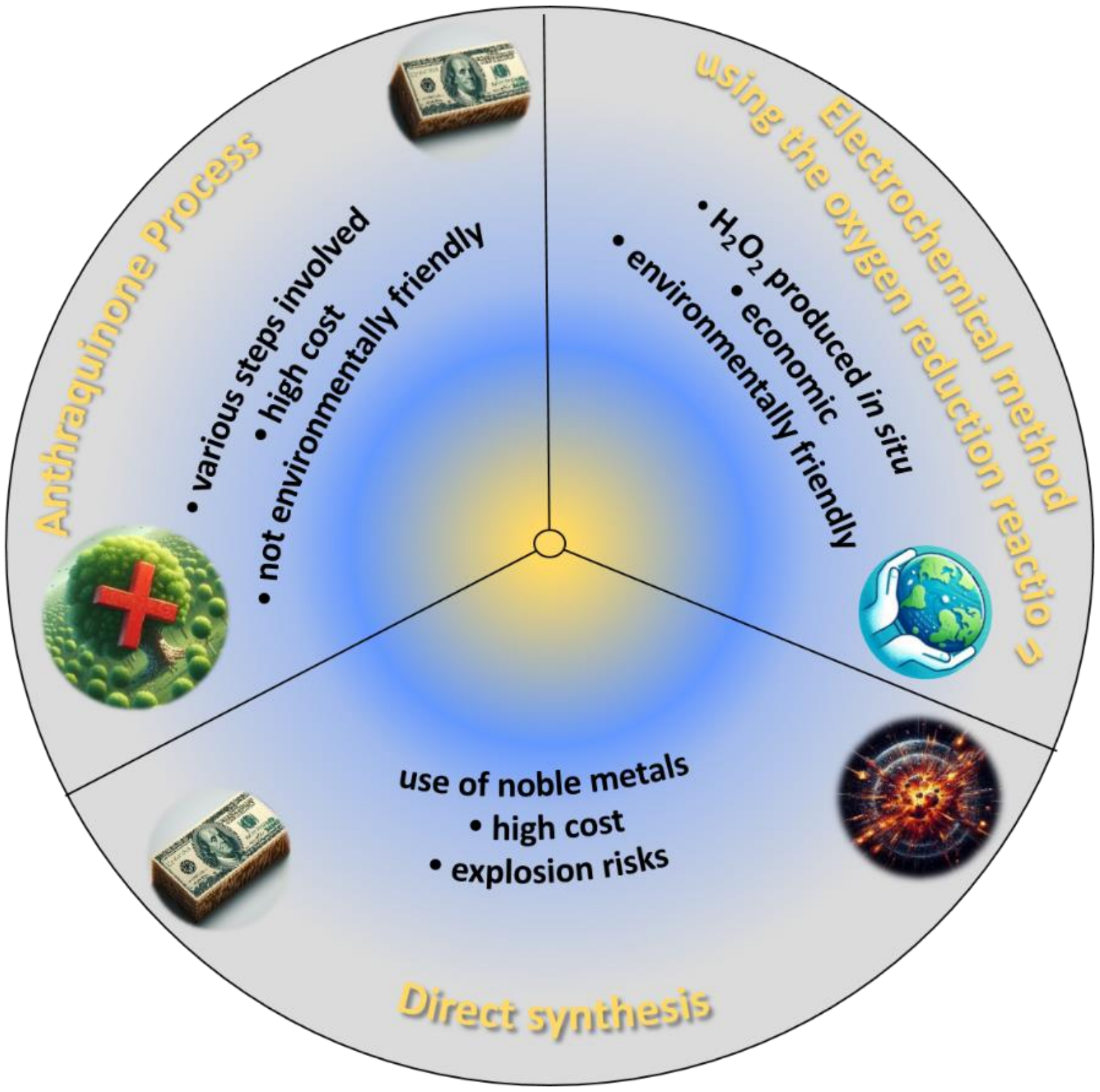

**Fig. 1**. Main methods of producing $H_2O_2$.

Fig. 2 presents the mechanism of ORR, which involves multiple reaction steps and intermediates, with the reaction pathway suggested by Wroblowa *et al*. (Wroblowa et al., 1976) being the most accepted. As shown in Fig. 2, the $O_2$ molecules adsorbed on the electrode surface can be reduced to $H_2O_2$ or $H_2O$ through the 2- or 4-electron ($e^-$) pathway, respectively. In the 2-electron pathway, $H_2O_2$ can still be further reduced to $H_2O$. The ORR can be applied in different areas, the 4-electron mechanism is employed in fuel cells and metal-air batteries, for example, and the 2-electron route, which is the focus of this review, can be applied in effluent treatment (Sires et al., 2014; Xia et al., 2015; Wang et al., 2020f).

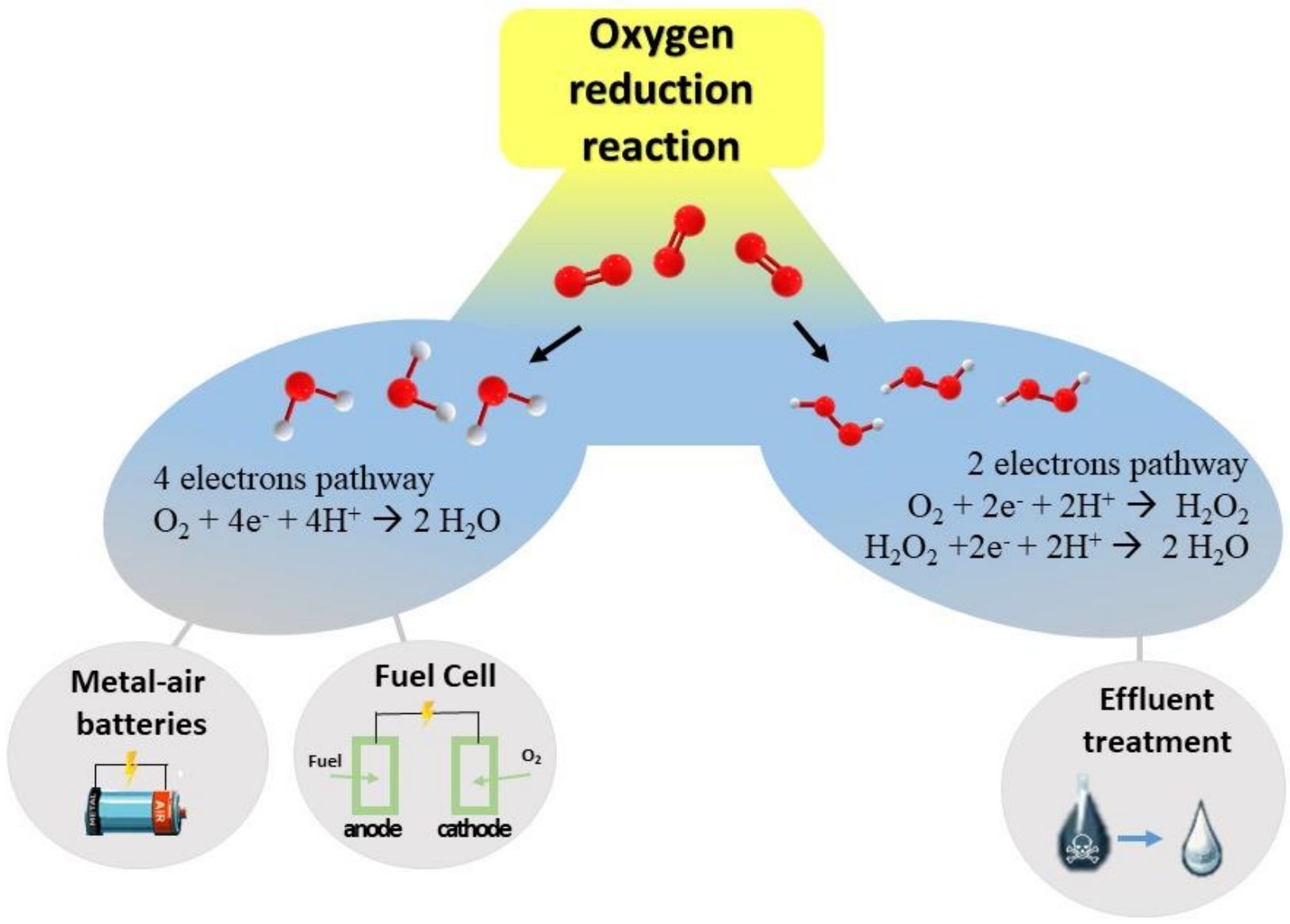

**Fig. 2**. Oxygen reduction reaction pathway.

Since ORR is a complex reaction, understanding how $O_2$ adsorption occurs on the electrode surface is essential to target the preferred pathway. As shown in Eq. (1), the $O_2$ adsorption on the electrode surface leads to the formation of $O_2$* species in an acid medium (the * symbol indicates that the species is adsorbed on the active site). Subsequently, the $O_2$* species is reduced by the action of an electron, forming the *OOH intermediate, as shown in Eq. (2). The *OOH intermediate is then reduced, forming $H_2O_2$ according to Eq. (3). Additionally, $H_2O_2$ or *OOH can also undergo further dissociation into O* and *OH leading to $H_2O$ formation, according to Eq. (4-6) (Wang, 2020; Santos et al., 2022).

$$O_2 + * \rightarrow O_2* \quad (1)$$

$$O_2* + H^+ + e^- \rightarrow *OOH \quad (2)$$

$$OOH* + H^+ + e^- \rightarrow H_2O_2* \quad (3)$$

$$*OOH \rightarrow *OH + O* \quad (4)$$

$$O* + H^+ + e^- \rightarrow *OH \quad (5)$$

$$OH* + H^+ + e^- \rightarrow H_2O* \quad (6)$$

The reaction pathway and the number of electrons that will be involved in the ORR are directly related to the oxygen molecule adsorption configuration. Fig. 3 shows the three ways that oxygen can adsorb on the electrode surface, namely the Griffith, Pauling, and bridge models. When adsorption occurs following the Griffith or bridge model, the ORR will follow the 4-electron pathway, forming $H_2O$. However, if adsorption follows the Pauling model, the ORR can occur via the 2-electron pathway, generating $H_2O_2$ [15]. The most efficient and selective electrocatalysts for the 2-electron ORR and $H_2O_2$ generation are noble metals. However, their high cost and scarcity make it impossible to generate $H_2O_2$ on a large scale (Siahrostami et al., 2013; Pizzutilo et al., 2017).

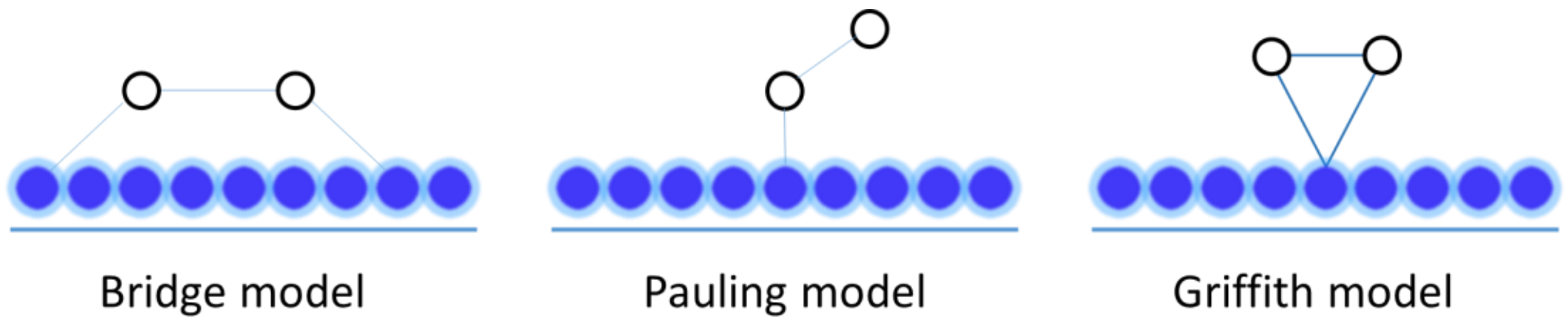

**Fig. 3.** Possible oxygen molecule adsorption models.

In this sense, carbon-based electrocatalysts have proven to be an attractive alternative for $H_2O_2$ synthesis. Since Berl *et al.* (Berl, 1939) reported for the first time the $H_2O_2$ production via ORR using an activated carbon cathode, carbon materials have been preferred over noble metals due to their low cost, abundance, and high selectivity for the 2-electron pathway (Assumpção et al., 2011; Iglesias et al., 2018; Kim et al., 2018; Wang et al., 2019; Cordeiro-Junior et al., 2020; Pena-Duarte et al., 2021; Santos et al., 2022). These advantages have caused different carbon materials to be heavily investigated, as well as strategies to improve their performance, resulting in numerous scientific articles. Therefore, many authors have dedicated themselves to summarizing advances in the area of $H_2O_2$ production via ORR using carbonaceous materials. Wei *et al*. (Wei et al., 2022a) presented, in a review, the main carbon-based electrocatalysts used in $H_2O_2$ electrogeneration, as well as the main modifications made to improve these electrocatalysts. Furthermore, this review reported some applications of electrocatalytic $H_2O_2$ production in environmental protection. In another review, Wang *et al*. (Wang et al., 2023b) summarized the main advances in the production of $H_2O_2$ via the 2-electron mechanism focusing on the influence of electrolyte pH, catalyst porous structure and selectivity. They also addressed the development of catalysts based on carbon derived from biomass and the main projects of electrochemical devices for the production of $H_2O_2$. Bu *et al*. (Bu et al., 2021) also addressed, in another review, the influence of the pH of the electrolyte and reactors for the production of $H_2O_2$. The review also brought the main strategies to optimize carbon-based materials for the production of $H_2O_2$. In another review, Li *et al*. (Li et al., 2022b) summarized the main studies on the role of hydrophilicity/hydrophobicity of carbon-based materials in relation to ORR and strategies to increase efficiency in $H_2O_2$ generation. Santos *et al.* (Santos et al., 2022) presented a review focused on the application of gas diffusion electrodes for $H_2O_2$ generation. Furthermore, they addressed the advances in the use of these electrodes on an industrial scale and their application in water and sewage treatment.

The present review focuses on individually exploring different carbon-based electrocatalysts (graphite, graphene, carbon nanotubes, and carbon black) used in the ORR for $H_2O_2$ electrogeneration. It also discusses the main approaches employed to enhance their performance. The review addresses the contributions of theoretical simulations in the field of $H_2O_2$ electrogeneration. Furthermore, stands out for presenting various advanced oxidative electrochemical processes utilizing carbon-based electrocatalysts, highlighting the radicals formed in each process.

Organized into sections, this review starts with Section 2, which presents different types of carbonaceous materials used for the 2-electron ORR, such as graphene, graphite, carbon nanotubes, and carbon black. Section 3 explores modifications made to improve the efficiency of $H_2O_2$ electrogeneration, including heteroatom-doped carbon, surface functionalization, and the insertion of metal oxides. Moving on to Section 4, recent theoretical simulations focused on the 2-electron ORR are presented. Section 5 continues the review by presenting advanced oxidative electrochemical processes such as electro-Fenton, photoelectro-Fenton, and solar electro-Fenton, highlighting the radicals formed in each process. Additionally, it reports the activity of carbon-based electrocatalysts in the degradation of various types of organic pollutants, such as pesticides, dyes, and pharmaceutical products. The review concludes with Section 6, providing Conclusions and a future outlook on the use of carbon-based materials in the electrogeneration of $H_2O_2$.

## 2. Carbon-based electrocatalysts

Carbon exhibits a substantial surface area coupled with exceptional electrical conductivity, leading to a highly advantageous scenario for the homogeneous distribution of catalytic sites and rapid charge transfer during the oxygen reduction reaction. This unique combination of properties significantly enhances the catalytic efficiency, making carbon a suitable material for this application (Muñoz-Morales et al., 2023).

Carbon-based materials have garnered widespread recognition for their outstanding performance as cathodes in the electrosynthesis of hydrogen peroxide, effectively serving as efficient alternatives to noble metal catalysts (Cordeiro-Junior et al., 2020; Zhang et al., 2020; An et al., 2022). The remarkable performance of carbon-based materials in $H_2O_2$ electrosynthesis can be attributed to a spectrum of exceptional properties. Derived from Earth's reserves, these materials boast abundant availability, establishing them as a sustainable and effortlessly obtainable resource. Their low cost makes them a cost-effective option, providing an economically viable solution for $H_2O_2$ electrosynthesis. With a non-toxic nature, these materials pose minimal environmental hazards, contributing to greener and safer chemical processes. The versatility and tunability of carbon-based materials allow tailoring their structures to suit specific applications, offering adaptability and efficiency. Impressive electrical conductivity facilitates swift charge transfer, leading to enhanced electrochemical performance. Furthermore, their inherent electrochemical stability ensures prolonged and reliable operation, a crucial factor for sustainable processes. Exhibiting an extensive surface area, carbon-based materials provide ample active centers for catalytic reactions, promoting efficient $H_2O_2$ production. Together, these inherent properties make carbon-based materials a promising choice for achieving superior performance in $H_2O_2$ electrosynthesis (Kim et al., 2018; Wang et al., 2020c; Zhou et al., 2021).

By capitalizing on these remarkable characteristics, carbon-based materials have carved out a crucial role in advancing the electrosynthesis of $H_2O_2$, paving the way for sustainable and eco-friendly chemical synthesis methodologies (Bu et al., 2021).

Indeed, a wide array of carbon-based materials has been successfully developed for $H_2O_2$ electrosynthesis. Among these, graphene, graphite, carbon nanotubes, and carbon black (Fig. 4) stand out as prominent examples, showcasing the versatility and effectiveness of carbon materials in this application. Through their unique properties and exceptional catalytic capabilities, these carbon-based materials have demonstrated great potential in advancing the electrosynthesis of $H_2O_2$, driving innovation in sustainable and efficient chemical synthesis methodologies (Marques Cordeiro-Junior et al., 2022).

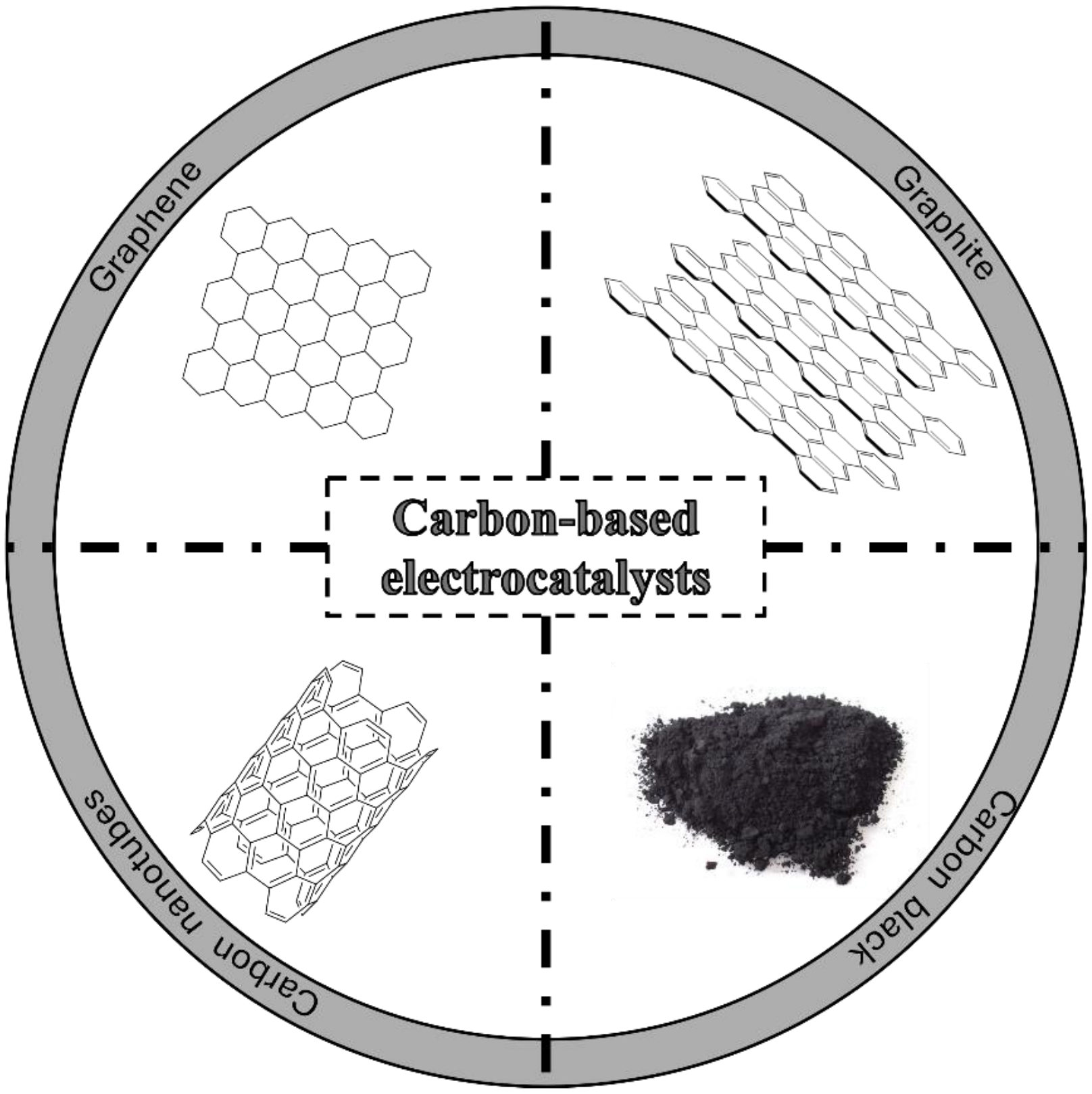

**Fig. 4.** Different types of carbon-based electrocatalysts.

### *2.1. Graphene*

Graphene is a unique two-dimensional material comprising $sp^2$ hybridized C atoms arranged in hexagonal rings, which stack together to form layered sheets. The commonly employed method for its synthesis involves the initial production of graphene oxide (GO), and then the reduction process that yields reduced graphene oxide (rGO) (Magne et al., 2022). GO and rGO composites have been the focus of several research works because of their diverse applications. Notably, they show great potential as electrocatalysts specifically for the 2-electron ORR, facilitating the $H_2O_2$ electrogeneration, as highlighted in a review by Y. Feng *et al.* (Feng et al., 2021).

Gao *et al.* (Gao et al., 2020) reported effective synthesis of graphene oxide through coating and annealing methods, yielding an impressive production of 6 mg $L^{-1}$ of $H_2O_2$ within a span of 60 minutes. This finding highlights the potential of GO as a catalyst for $H_2O_2$ generation. In the same year, a review paper (Yu and Breslin, 2020) underscored the potential applications of 2D graphene and graphene-like for $H_2O_2$ generation. The review brought attention to the versatility and exciting possibilities of these advanced materials for efficient $H_2O_2$ production, emphasizing their potential impact on various industries and applications.

In a paper published in 2022, Li *et al.* (Li et al., 2022a) demonstrated the production capabilities of rGO synthetic fabric when reduced with HI and $NaBH_4$. The reduction with HI resulted in an estimated $H_2O_2$ production of 4.78 mg $h^{-1}$ $cm^{-2}$, with a current efficiency of 57.97% within 60 min. Conversely, the reduction with $NaBH_4$ exhibited a higher estimated $H_2O_2$ production of 5.59 mg $h^{-1}$ $cm^{-2}$, with a current efficiency of 63.94%.

A research work carried out by Li *et al.* (Li et al., 2022c) observed that varying the reducing temperature of graphene oxide had a significant impact on the $H_2O_2$ generation rate, resulting in an impressive rate of 20.4 mg $h^{-1}$ $cm^{-2}$. Interestingly, while the controlled reduction process led to higher $H_2O_2$ production, unreduced graphene demonstrated the highest peroxide selectivity.

Furthermore, a work by Lee *et al.* in 2022 (Lee et al., 2022) showcased the remarkable potential of crumpled graphene electrocatalysts, with the presence of highly active defect arrangements and oxygen-functionalized groups. This innovative catalyst exhibited exceptional selectivity to produce $H_2O_2$, achieving an impressive rate of 473.9 mmol $gcat^{-1}$ $h^{-1}$, and displayed remarkable stability over a testing period of 46 hours. These discoveries provide fresh insights into the promising application of crumpled graphene in the field of electrocatalysis for $H_2O_2$ synthesis.

### *2.2. Graphite*

Graphite felt (GF) represents a standard three-dimensional electrode possessing remarkable electrical conductivity (370.37 S $m^{-1}$). This exceptional conductivity has rendered it widely utilized for hydrogen peroxide electrogeneration and electrochemical advanced oxidation processes (EAOP) aimed at pollutant degradation. Notably, the extensive surface area and robust mechanical properties of graphite felt make it an excellent choice as a cathode for the $H_2O_2$ electrogeneration. These advantageous characteristics contribute to its effectiveness in generating $H_2O_2$, enabling its applications for water treatment and pollutant removal (Miao et al., 2014; Zhao et al., 2022).

In a significant breakthrough in 2019, W. Zhou *et al.* (Zhou et al., 2019b) published their findings, introducing an environmentally friendly and pragmatic electrochemical modification technique for graphite felt. This method enables *in situ* operation and facilitates sequential cathode modification, leading to hydrogen peroxide generation. Remarkably, the modified graphite felt demonstrated considerable improvement, achieving an impressive 183.3% higher $H_2O_2$ yield compared to conventional methods. Furthermore, the researchers conducted long-term stability testing over 30 cycles, and the modified graphite exhibited excellent durability and consistency throughout the testing period.

In 2020, Diouf *et al.* (Diouf et al., 2020) published a study on the $H_2O_2$ electrogeneration. They employed a graphite cathode extracted from exhausted batteries for this purpose. The study extensively explored hydrogen peroxide production parameters, including the characteristics and concentration of the electrolyte, pH levels, the presence of $Fe^{2+}$ ions, and oxygen injection. The researchers made a comparative analysis by evaluating the performance against a vitreous carbon electrode. This research represents a significant step towards sustainable and eco-friendly methodologies for $H_2O_2$ generation, with the potential to repurpose used materials for beneficial applications. The findings shed light on the viability of graphite cathodes from exhausted batteries as efficient alternatives, contributing to the growing body of knowledge on green electrochemical processes and their impact on the environment.

Also in 2020, J. Zhou *et al.* (Zhou et al., 2020) introduced a remarkable integrated strategy aimed at modulating the structure of graphitic felts. The researchers conducted a thorough investigation of the various treatment processes and their significant contributions to reaction activity. Through chemical and electrochemical oxidation, the team successfully induced the creation of imperfections in the structure and the introduction of oxygen-based functional moieties. These modifications significantly contributed to improving the reaction kinetics, ultimately leading to enhanced $H_2O_2$ production rates. Additionally, thermal treatment was employed to facilitate the efficient interfacial transfer of $O_2$ during the 2-electron ORR. The combination of these treatment methods in a synergistic approach resulted in an impressive five-fold increase in the $H_2O_2$ yield. This substantial improvement demonstrates the effectiveness of their integrated strategy in maximizing $H_2O_2$ synthesis. The study's findings offer valuable insights into the manipulation of graphitic felt structures for advanced electrochemical applications.

In 2021, Wang and Lin (Wang and Lin, 2021) conducted a study on the utilization of graphite felt and its modifications for $H_2O_2$ electrogeneration and electro-Fenton applications. Their research yielded a notable achievement of approximately 5 mg $L^{-1}$ of peroxide generation using pure graphite felt. In the subsequent year, K. Wang *et al.* (Wang et al., 2022a) investigated the impact of several parameters on the $H_2O_2$ electrogeneration using graphite powder. Overall, these two studies contributed with valuable insights for the optimization of graphite-based materials for efficient $H_2O_2$ generation, paving the way for advancements in electrochemical applications and water treatment processes.

In 2022, Xu *et al.* (Xu et al., 2022a) presented a noteworthy study investigating the potential of graphite felt anodized with sodium hydroxide, ammonium bicarbonate, or sulfuric acid aqueous solutions as cathodes for *in situ* hydrogen peroxide production. The researchers performed electrolysis at -0.60 V (vs. SCE) for 120 min and obtained promising results. Among the different anodization methods, graphite felt anodized with 0.2 M $H_2SO_4$ exhibited exceptional performance, achieving an $H_2O_2$ yield of up to 110.5 mg $L^{-1}$ in a 0.05 M $Na_2SO_4$ electrolyte. When compared to the raw graphite felt used as a cathode under the same conditions, this anodization process led to a remarkable 15.85-fold increase in the $H_2O_2$ yield. Further exploration and optimization of these anodization techniques could pave the way for more sustainable electrochemical processes in the future.

### *2.3. Carbon nanotubes*

Carbon nanotubes (CNTs) are known for their remarkable properties that make them very promising electrode materials. They exhibit a range of unique characteristics, including remarkable electrical conductivity, large surface area, cost-effectiveness, as well as easy and precisely tunable atomic arrangement via processes like heteroatom doping and surface functionalization. Additionally, CNTs are considerably stable even under harsh reaction conditions, while maintaining significant mechanical strength.

These exceptional attributes make CNTs stand out as electrode materials in various applications. Their high electrical conductivity promotes efficient electron transfer, which is pivotal for enhancing their performance for energy storage and conversion applications. Furthermore, CNTs' properties can be tuned at an atomic level to customize the structures to meet specific application requirements. Moreover, the substantial surface area of CNTs implies a higher amount of active sites, leading to increased electrochemical activity. This aspect is particularly advantageous for applications in batteries, supercapacitors, and fuel cells, where fast charge-discharge rates and high-power output are desirable. Besides their exceptional electrical properties, the mechanical strength of CNTs provides enhanced durability, ensuring long-term stability and reliability in demanding environments (van Dommele et al., 2006; Zhang et al., 2008; Gong et al., 2009).

In 2019, Xia *et al.* (Xia et al., 2019) published fascinating findings regarding the electrogeneration of hydrogen peroxide $H_2O_2$ utilizing both pure and phosphorus-doped carbon nanotubes. Their research shed light on the electrochemical properties and potential applications of these nanotubes in $H_2O_2$ synthesis. Subsequently, in 2020, G. Pan, X. Sun, and Z. Sun (Pan et al., 2020) made an exciting breakthrough by introducing a highly efficient cathode for $H_2O_2$ electrogeneration. They developed a co-modified graphite felt electrode using a combination of multi-walled carbon nanotubes (MWCNTs) and carbon black (CB). The resulting electrode, labeled MWCNTs-CB/GF, demonstrated exceptional performance for the $H_2O_2$ production. After 120 minutes of operation, the MWCNTs-CB/GF electrode achieved an $H_2O_2$ generation of 309.0 mg $L^{-1}$ with an impressive current efficiency of 60.9%. This outstanding performance highlighted the MWCNTs-CB/GF electrode potential as a superior catalyst for electrochemical $H_2O_2$ synthesis.

Also in 2020, a study conducted by L. Tao, Y. Yang, and F. Yu (Tao et al., 2020) introduced a novel modified electrode for cathodic applications by incorporating active carbon fibers (ACFs) with porous carbon (PC) and CNTs. The researchers investigated the influence of different CNT mass ratios on both $H_2O$ production and electric energy consumption. Remarkably, they found the best ratio was 1:7. The study revealed a significant enhancement in $H_2O_2$ production when using PC-CNTs/ACFs as the cathode, with an impressive output of 1,554.55 mg $L^{-1}$. In stark contrast, the $H_2O_2$ production with raw ACFs as the cathode was substantially lower at just 59.96 mg $L^{-1}$. This striking difference highlights the exceptional performance and efficacy of the novel modified electrode.

Wang *et al.* (Wang et al., 2020b) conducted a study utilizing oxidized carbon nanotubes as catalysts to emphasize the role of $O_2$ supply in such processes. They carried out a comprehensive evaluation by comparing the current efficiency of hydrogen peroxide generation on two distinct electrode setups: a rotating ring-disk electrode (RRDE) and a gas diffusion electrode (GDE). By utilizing oxidized carbon nanotubes as catalysts, the researchers aimed to elucidate the importance of oxygen availability in $H_2O_2$ generation. The comparison between the RRDE and the GDE allowed them to precisely assess the impact of different oxygen supply mechanisms on the efficiency of $H_2O_2$ production. Their findings provided useful perspectives on the intricacies of the electrochemical process and how oxygen availability influences the overall performance of the catalyst. Understanding the correlation between oxygen supply and $H_2O_2$ generation is crucial for optimizing and enhancing the efficiency of such processes.

In a study published in 2022 (Liu et al., 2022a), Liu *et al.* showcased the impressive potential of optimized oxygen-functionalized CNTs (O-CNTs). These enhanced O-CNTs exhibited an exceptional selectivity of approximately 92% for producing $H_2O_2$ across a broad voltage range in a 0.1 mol $L^{-1}$ KOH solution. Additionally, the study revealed an astonishingly high hydrogen peroxide production rate of 296.84 mmol $L^{-1}$ $g^{-1}$ cat $h^{-1}$, emphasizing the remarkable efficiency of these nanotubes as catalysts for $H_2O_2$ synthesis.

### *2.4. Carbon black*

Carbon black is a type of nanocarbon consisting of spherical particles primarily composed of nearly pure elemental carbon. This ultralight and extremely fine black powder exhibits a density ranging from 1.70 to 1.90 g $cm^{-3}$. It is produced through the incomplete combustion or thermal decomposition of carbon-containing substances, such as coal, natural gas, heavy oil, and fuel oil, under limited air supply conditions. Carbon black has attracted global attention thanks to its intricate structure, fundamental properties, and diverse applications. Particularly in catalysis, carbon black has become popular due to its exceptional electrical conductivity, high surface area, and remarkable stability (Moraes et al., 2015; Zeng et al., 2017; Gautam and Verma, 2019). Similar to other carbon-based materials presented in this section, carbon black's high surface area provides many active sites for catalytic reactions and its stability ensures long-lasting performance, even under challenging catalytic conditions. The unique combination of these properties makes carbon black an interesting material for several applications, such as environmental remediation, energy conversion, and industrial chemical synthesis. The ongoing research and exploration of carbon black's capabilities continue to expand its potential for addressing global challenges and advancing catalytic science (Moraes et al., 2015; Zeng et al., 2017; Gautam and Verma, 2019).

Printex L6 carbon (PL6C) stands out as one of the most extensively employed carbon black in the $H_2O_2$ electrogeneration. Over the last few years, extensive research has been conducted on PL6C, primarily focusing on surface modifications to enhance its performance (Cordeiro-Junior et al., 2020; Trevelin et al., 2020; Trevelin et al., 2023). In its pure form, PL6C exhibits an impressive $H_2O_2$ selectivity surpassing 80%, making it an attractive catalyst for this electrochemical process. Furthermore, it is a stable and lasting catalyst, with a maximum lifetime of 36 hours or approximately 7.5 days of uninterrupted operation. Within this time frame, it consistently produces a range between 150 to 350 mg $L^{-1}$ of hydrogen peroxide, holding great promise for applications in various industries (Cordeiro-Junior et al., 2022; Cordeiro Junior et al., 2022; Marques Cordeiro-Junior et al., 2022; O. Silva et al., 2022; Kronka et al., 2023). As investigations into PL6C continue, it is evident that its role as a key catalyst in $H_2O_2$ production will expand, driving innovative solutions for sustainable and efficient electrogeneration processes.

Vulcan XC-72 is another well-known type of carbon black specifically designed to have a high surface area and porosity. Vulcan XC-72 can effectively promote several catalytic reactions, including those related to $H_2O_2$ generation (Trevelin et al., 2023). In electrochemical processes, it works as a highly efficient catalyst support material (Assumpção et al., 2011; Antonin et al., 2017). It is worth noting that researchers continue to explore different catalyst materials and electrochemical setups to improve efficiency and scalability and, in this context, Vulcan XC-72 carbon is a very promising candidate (Pérez-Rodríguez et al., 2018). Recently published studies have showcased its remarkable potential for $H_2O_2$ generation (Aveiro et al., 2018a; Aveiro et al., 2018b; Paz et al., 2018; Pinheiro et al., 2019; Kornienko et al., 2020; Paz et al., 2020; Machado et al., 2022; Moura et al., 2023). Further surface modifications can also be applied to Vulcan XC 72 to improve its catalytic performance for this application (Aveiro et al., 2018a; Aveiro et al., 2018b; Paz et al., 2018; Pinheiro et al., 2019; Kornienko et al., 2020; Paz et al., 2020; Machado et al., 2022; Moura et al., 2023).

The subsequent section of this review delves into a more comprehensive examination of surface modifications and diverse catalyst types seamlessly integrated into carbonaceous matrices. Despite the inherent goodness of carbon-based materials in facilitating the 2-electron ORR, it is acknowledged that they may not be optimal electrocatalysts for this process. Therefore, the continuation of the text addresses the various modifications undertaken to optimize their performance. These advancements are strategically designed to elevate both efficiency and catalytic activity, ultimately bolstering the sustainable production of hydrogen peroxide through electrochemical means.

## 3. Increased $H_2O_2$ electrogeneration through modification of carbonaceous materials

The quest for efficient and sustainable $H_2O_2$ electrogeneration has driven significant research towards enhancing the performance of carbon-based catalysts. Tremendous efforts have been made by the scientific community to explore various approaches, including hetero-atom doping, surface functionalization, and catalyst engineering to modify carbonaceous materials (diagrammed in Fig. 5). These modifications introduce defects in the carbon structure and functional groups on the catalyst surface, improving the selectivity for $H_2O_2$ production via the 2-electron ORR. This section presents novel strategies and the achieved advancements in improving $H_2O_2$ electrogeneration through the modification of carbonaceous materials.

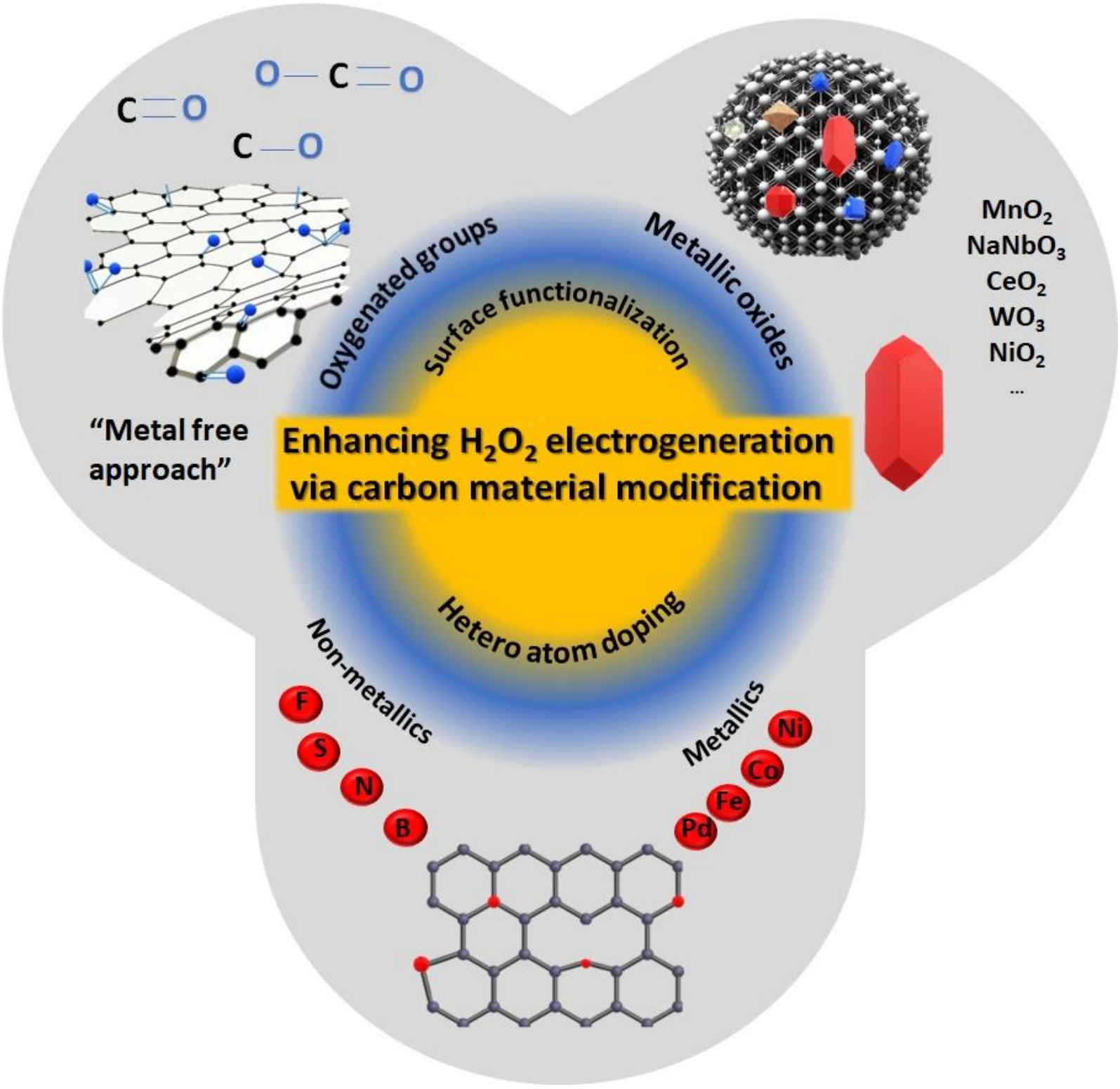

**Fig. 5**. Some approaches used to increase $H_2O_2$ electrogeneration by carbonaceous materials.

### *3.1. Doped carbon materials*

#### 3.1.1. Hetero-atom doped carbon materials

Among the possible heteroatoms employed for doping carbon materials, nitrogen is the most extensively reported. Despite the evident performance enhancements observed in nitrogen-doped carbon catalysts, the true nature of catalytically active sites has been largely overlooked, primarily due to the co-existence of several nitrogen functional groups within nitrogen-doped materials.

N-doped carbon materials can be synthesized through various methods, each tailored to achieve specific properties and applications. Common techniques include (i) Chemical vapor deposition (CVD): in this method, a carbon precursor is exposed to a nitrogen-containing gas at high temperatures, resulting in the incorporation of nitrogen atoms into the growing carbon structure; (ii) pyrolysis of nitrogen-containing precursors: nitrogen-containing organic compounds, such as amines or nitriles, are pyrolyzed in an inert atmosphere to produce N-doped carbon materials; (iii) template-assisted synthesis: a template such as a zeolite or silica, is impregnated with a carbon precursor and a nitrogen source. After carbonization, the template is removed, leaving behind N-doped carbon with controlled porosity; (iv) electrochemical methods: these can be employed to introduce nitrogen into carbon materials, for example, during the electroreduction of nitrogen-containing electrolytes; (v) physical mixing with nitrogen-containing substances, such as urea or ammonia, followed by heat treatment to promote N-doping (Kamedulski et al., 2022; Zhang and Waki, 2022). Each synthesis method offers unique advantages and can be optimized to tailor the N-doped carbon materials for specific applications, such as energy storage, catalysis, and environmental remediation.

In N-doped carbon catalysts for the ORR, different types of nitrogen-containing functional groups can work as active sites that will contribute to the general catalytic performance. These functional groups can promote the ORR and improve the efficiency of carbon-based materials. Some of the key nitrogen-containing functional groups found in N-doped carbon catalysts for ORR are: pyridinic nitrogen (Pyri-N), pyrrolic nitrogen (Pyrr-N), graphitic nitrogen (Grap-N), quaternary and pyridinic-N-oxide. Pyridinic nitrogen (Pyri-N) species, known for their ability to influence nearby carbon atoms and topological carbon defects, are typically recognized as active sites. Several studies have suggested that N-doped carbon-based catalysts tend to promote the four-electron ORR pathway. However, Wan and collaborators (Wan et al., 2022) investigated the effect of N-doping on the 2-$e^-$ ORR pathway. The authors reported well-controlled N-doping carbon nanotube and graphene using a microwave-assisted pulse heating method, obtaining Grap-N, Pyrr-N, and Pyri-N configurations with high centrality. Their findings showed that the Pyri-N sites were the most effective for the 2e- ORR, with high $O_2$-to-$H_2O_2$ conversion, and a selectivity of 93.5% on N-CNT and 98.35 on N-rGO. These results illustrate the importance of fine-tuning the catalytic sites at an atomic level as a strategy to change the material selectivity towards the preferred reaction pathway.

Starting from different commercially available carbon black, Kamedulski *et al.* (Kamedulski et al., 2022) investigated N-doping promoted by gamma radiation treatment. Their results showed that the carbon black source is crucial for the catalytic performance and selectivity. Overall, N-doping improved the selectivity towards the 2-$e^-$ pathway, but the commercial carbon black PK 1-3 Norit exhibited the best activity among the selected materials after that functionalization. N-doped PK 1-3 Norit yielded a 1.95 electron transfer in ORR, which means a high $O_2$-to-$H_2O_2$ conversion rate, in comparison with 2.21 on pure PK 1-3 Norit.

Recent works also report N and O co-doping as an approach to improve the 2-electron selectivity of carbon-based materials (Koh et al., 2022; Xu et al., 2022c; Zhang et al., 2022a; Sun et al., 2023b). Zhang *et al*. (Zhang et al., 2022a) reported a catalyst based on pyrolyzed biomass, co-doped with N and O functional groups. The obtained catalyst had a 4 times higher surface area in comparison with the undoped material. The presence of N and O functional groups on the surface resulted in a high selectivity of 90% for the 2-electron ORR across a broad potential range, with a faradaic efficiency of about 80%.

Other non-metallic dopants have also demonstrated potential as catalysts for hydrogen peroxide electrosynthesis. The introduction of dopants such as sulfur (Xu et al., 2022b; Qian et al., 2023; Wang et al., 2023a), boron (Liu et al., 2022b; Fan et al., 2023a), and fluorine (Zeng et al., 2022) can induce structural changes and alter the surface properties of carbon, influencing the kinetics and efficiency of the 2-electron pathway. In this sense, Xu *et al.* (Xu et al., 2022b) synthesized an S, N-co-doped few-layered graphene that exhibited outstanding activity for the 2-electron ORR, with $H_2O_2$ selectivity reaching an impressive range of 90%–100%. The calculated faradaic efficiency achieved a maximum value of 93%. The observed results indicate that the combined presence of oxidized sulfur and pyridinic N had a positive impact on the catalyst's performance, reducing the Fermi level of the active edge carbon sites. Furthermore, the *OOH intermediate binding energy optimization favored the 2-electron pathway, contributing to the exceptional selectivity and efficiency for $H_2O_2$ production.

The carbon doping with heteroatoms which sizes and electronegativities distinct from those of carbon atoms into the carbon host has the potential to lead charge redistribution among carbon atoms through charge transfer with the heteroatoms. This, in turn, results in the modification of geometric, electronic properties causing changes in other characteristics, such as morphologies, enhancing porosity, and additional actives functional groups (Ma et al.; Hu and Dai, 2019). Wang *et al*.(Wang et al., 2023a) results demonstrate that S and N doping can positively alter the electronic nature of carbon atoms adjacent to heteroatoms. This alteration is beneficial for oxygen adsorption, leading to increased selectivity for the ORR and $H_2O_2$ yield. However, in Juan Wan studies (Wan et al., 2022) the investigation of N-doped carbon nanotubes (CNTs) found contrasting results. N-doping increased the ORR activity but decreased the selectivity for ORR 2e-, these findings were deemed optimal for other catalyst-ORR applications, such as fuel cell devices. In conclusion, one of the main features governing the tuning of doping in carbon for ORR selectivity towards hydrogen peroxide is the nature of the formed active site and the specific interactions with $O_2$ molecules.

Doping carbon-based materials with metal atoms is another interesting strategy to tune the catalytic activity. The synergistic effect between metal and carbon defective sites can enhance the ORR and $H_2O_2$ selectivity. This is particularly attractive when non-noble metals are employed, given their relative abundance and low cost. Wang *et al.* (Wang et al., 2022b) reported a high-performance electrocatalyst featuring single-atom nitrogen-coordinated Pd, synthesized via one-step thermolysis of Pd-doped zeolitic imidazolate frameworks (ZIFs). The obtained catalyst exhibited well-dispersed Pd-N-C and optimal performance for the 2-electron ORR. The achieved $H_2O_2$ selectivity was above 95%, in comparison with the Pd-free catalyst with ~50%. Acknowledging that the selectivity was related to Pd sites, the authors carried out density functional theory (DFT) calculations to reveal that the Pd-$N_4$ catalytic sites have a thermodynamic preference for breaking the *-O bond rather than the O-O bond, which accounts for the high selectivity for $H_2O_2$ production. Despite the use of a noble metal with a high associated cost, this work shows that small amounts of dopant already resulted in a great impact on the catalytic activity.

Over the past few years, co-doping with transition metals and nitrogen has become the focus of various research works. Fe/N co-doped graphene with a high efficiency for 2-electron ORR was described by Yu *et al.* (Yu et al., 2023). The enhancement of the ORR performance was attributed to the strong interaction between N-Fe bonds. This interaction between FeOx and N accelerates electron transfer during the electrochemical reaction, thereby boosting the $H_2O_2$ generation. This material achieved 92.7% $H_2O_2$ selectivity, while the undoped catalyst reached 51.1 % $H_2O_2$ selectivity. Furthermore, the Fe-doping has a bifunctional role acting as a Fenton-like site for hydroxyl radicals *in situ* generation. López *et al.* (Barranco-Lopez et al., 2023) investigated Fe-doped carbon xerogels, synthesized by a one-step sol-gel polymerization, as bifunctional electro-Fenton catalysts and noted that the material facilitated both the *in situ* $H_2O_2$ production and its conversion into hydroxyl radicals. In this case, Fe significantly enhances the electrocatalytic activity and Fe-O-C active sites promote mainly the 2-electron pathway with high selectivity.

The selectivity of a catalyst for the 2-electron ORR primarily depends on how the *OOH intermediate absorbs on the active sites. This absorption behavior, including its strength and configuration, is significantly influenced by the catalyst's electronic structure, specifically its metal d-band center (Liu et al., 2023). Thus, carbon electrocatalysts functionalized with transition metals are among the most promising and cost-effective alternatives for catalyzing this reaction. In this sense, Hao Hu and collaborators (Hu et al., 2023a) systematically investigated different carbon-supported transition metal (Mn, Fe, Co, Ni, Cu) catalysts. The materials were synthesized using a straightforward pyrolysis method mediated by zinc. Different transition metals led to significant variations in the oxygen content within the M-C catalysts. The results indicated that Cu-C and Co-C are promising catalysts for the ORR, mainly through the 4-electron mechanism. On the other hand, Fe-C and Mn-C catalysts exhibit a combination of both 2 and 4-electron pathways. Additionally, Ni-C works as an efficient catalyst for $H_2O_2$ generation, following the 2-electron pathway. The research reveals a correlation between the ORR selectivity of M-C catalysts and the content and type of oxygen-containing functional groups within the structures. The authors suggest that C-O-C functional groups are probably active sites for the 4-electron pathway.

Some reports show Pd-based modified carbon as a promising catalyst for efficient 2-electron ORR (Fortunato et al., 2020; Fortunato et al., 2022; Cordeiro Junior et al., 2023). In this sense, Fortunato (Fortunato et al., 2020) developed a catalyst based on less than 1 wt% of Pd nanoparticles dispersed on Printex L6 (PL6C) carbon black prepared by a simple hydrothermal method. The authors achieved a carbon catalyst with well-dispersed Pd sites with an improved electrochemical active surface and an 84 % $H_2O_2$ yield. When gas diffusion electrodes based on 1% Pd/ PL6C were used, a considerable improvement in the process faradaic efficiency and a considerable reduction in energy consumption was observed for the $H_2O_2$ electrogeneration compared to pure PL6C. Another work in this direction has been carried out by Cordeiro Junior *et al.* (Cordeiro Junior et al., 2023). In this work, a Pd-based complex/PL6C catalyst shifted the ORR onset potential to more positive values and achieved a remarkable 97% selectivity toward $H_2O_2$ production. This electrocatalytic effect was attributed to π-π interactions between the $Pd^{2+}$ $dz^2$ orbital and nearby carbon atoms with oxygenated functional groups within the carbonaceous matrix. The theoretical analyses indicated that $Pd^{2+}$ in combination with oxygenated carboxyl functional groups, denoted as COOH, played a pivotal role in facilitating the interaction between $O_2$ and hydronium ions to generate the OOH* species. Overall, the Pd-based complex/PL6C catalyst is a promising candidate for efficient and selective hydrogen peroxide production.

Finally, the literature indicates that carbon materials functionalized with metal atoms can be highly effective for the 4-electron ORR. Therefore, the selection and the synthesis methodology of the catalyst play a crucial role in determining the ORR mechanism selectivity, tuning the catalyst surface with specific active sites for 2-electron ORR.

### *3.2. Surface functionalization*

#### 3.2.1. Oxygenated groups

Carbon materials containing oxygen functional groups have shown great potential as efficient electrocatalysts for the 2-electron ORR. The synthesis of these materials generally follows two main approaches: top-down and bottom-up. In the top-down approach, oxygen functional groups are introduced into pre-existing carbon materials through oxidative treatments like $HNO_3$, plasma, and heat treatments. These treatments lead to the implantation of oxygen functionalities into the carbon matrix, enhancing its catalytic activity for ORR. On the other hand, the bottom-up approach involves the transformation of oxygen- and carbon-containing precursors via pyrolysis at high temperatures. This process results in the formation of graphitized carbon materials with inherent oxygen functionalities. Both approaches offer distinct advantages, and researchers can tailor the synthesis method based on specific applications and desired properties. O-doped carbon materials hold immense potential for advancing sustainable energy conversion technologies due to their metal-free nature and high catalytic activity (Lee et al., 2023).

Potential development trends for electrochemical $H_2O_2$ production include the identification of catalytic active sites and the focusing modification of the carbon with active species. These approaches aim to enhance the efficiency and selectivity of $H_2O_2$ electrogeneration, paving the way for more sustainable and efficient electrochemical processes. Li *et al.* (Li et al., 2022a) conducted a study involving a 3D reduced graphene oxide synthetic fabric (rGOSF) cathode, aiming to achieve a high electrocatalytic activity for $H_2O_2$ generation. This investigation incorporated a multi-step reduction approach designed to fine-tune the oxygen functional groups present on the rGOSF surface. Subsequent electrochemical assessments indicated that the active sites responsible for $H_2O_2$ generation were primarily derived from carboxyl groups. Density functional theory (DFT) calculations further confirmed the crucial role of carboxyl groups in the oxygen reduction process to generate $H_2O_2$. A carefully chosen carboxyl-rich functional molecule was employed to develop an active species-modified rGOSF cathode through a precise wet co-spinning assembly. The resulting carboxyl-functionalized rGOSF exhibited significantly enhanced activity for $H_2O_2$ generation with 90 % $H_2O_2$ selectivity and a current efficiency of 63.9 %.

Investigating environmentally friendly processes, Fan and collaborators (Fan et al., 2022) reported a biomass-based graphene with tunable oxygen species fabricated by a $CO_2$ laser. The laser-induced graphene (LIG) demonstrated outstanding catalytic performance, boasting an impressive selectivity of over 85% of $O_2$-to-$H_2O_2$ ORR, along with a high Faraday efficiency surpassing 78% and a mass activity of 814 mmol $g_{catalyst}^{-1}$ $h^{-1}$ using a flow cell setup. These remarkable results surpass the performance of most reported carbon-based electrocatalysts. DFT analyses indicate that the meta-C atoms adjacent to C-O and O=C=O species play a pivotal role as key catalytic sites, further contributing to the superior performance of the LIG catalyst. Overall, functionalizing carbon with oxygenated groups is a very promising direction for optimizing catalytic materials for the $H_2O_2$ electrogeneration (Chu et al., 2022; Lee et al., 2023).

In a recent study, Chu *et al.* (Chu et al., 2022) developed a novel electrocatalyst by covalently linking anthraquinone to amino-functionalized carbon nanotubes (NCNT-AQ) and modifying it with PTFE on carbon felt. The introduction of anthraquinone through chemical bonding significantly improved the current efficiency of the electrocatalyst. After 1 h of operation, the current efficiency of CNT, NCNT, and NCNT-AQ reached 57.2%, 54.5%, and an impressive 89.2%, respectively. Furthermore, the $H_2O_2$ generation rate achieved an outstanding value of 2.09 M $g_{catalyst}^{-1}$ $h^{-1}$ at a current density of 50 mA $cm^{-2}$, which was 4.45 times higher than that achieved by traditional electrodes. These results highlight the tremendous potential of the NCNT-AQ electrocatalyst for efficient $H_2O_2$ production, surpassing conventional methods by a significant margin.

Briefly, the literature shows the strong effort from the scientific community to explore and develop increasingly efficient modifiers and catalytic materials. These advancements highlight the ongoing dedication to finding innovative and environmentally friendly processes, with the potential to revolutionize $H_2O_2$ production and other electrochemical applications.

### 3.2.2. Metallic oxides

Recent research works demonstrated that metal oxides also have good electrocatalytic activity for $H_2O_2$ production. Metal oxides have drawn attention as promising modifying-carbon materials due to their low cost and relative abundance, in contrast to noble metals. These properties facilitate metal oxides applications on a large scale. Notably, metal oxides' morphological and electronic properties can be tuned by changing simple parameters during the synthesis process, making them very promising catalysts.

Recently, several metal oxides have been reported as a strategy to modify carbon electrocatalysts and improve their activity for $H_2O_2$ electrogeneration. Some examples of carbon-based materials functionalized with metal oxides include $Nb_2O_5$ (Valim et al., 2021; Trench et al., 2023; Trevelin et al., 2023), $RuO_5$ (Valim et al., 2021), $NiO_2$ (Wu et al., 2022; Nosan et al., 2023), $MnO_2$ (Aveiro et al., 2018a; Moura et al., 2023), $ZrO_2$ (Kronka et al., 2023), $TiO_2$ (Tu et al., 2022), $WO_3$ (Paz et al., 2020; Xu et al., 2023b), $CeO_2$ (Pinheiro et al., 2018), and $NaNbO_3$ (Antonin et al., 2023). In a nutshell, the modification of carbon with oxides can significantly enhance the electrogeneration of $H_2O_2$ through the ORR by various mechanisms. The addition of oxides on the carbon surface creates new active sites, facilitating the electrochemical adsorption of oxygen and hydronium ions ($H^+$) during the ORR. This favorable interaction promotes the formation of $H_2O_2$ as an intermediate product. Additionally, certain oxides can improve the electrical conductivity (and for instance the hydrophilicity) of the modified carbon material, leading to more efficient electron transfer during the ORR. This enhanced conductivity contributes to a higher electrochemical reaction rate and improved hydrogen peroxide generation. Lastly, specific oxides can act as regulators of active sites on the carbon surface. By influencing the structure and geometry of carbon atoms at the surface, they impact the selectivity for $H_2O_2$ formation during the ORR, making the process more efficient and selective.

In this sense, Antonin *et al.* (Antonin et al., 2023) conducted a study to explore the catalytic activity of $NaNbO_3$ microcubes decorated with $CeO_2$ nanorods on carbon. Their findings revealed a significant improvement in $H_2O_2$ electrogeneration when using $NaNbO_3$@$CeO_2$/C-based materials compared to pure Vulcan XC 72. Notably, the 1% $NaNbO_3$@$CeO_2$/C electrocatalyst exhibited a lower initial potential for the ORR, favoring the 2-electron mechanism with an impressive 82% selectivity towards $H_2O_2$. The introduction of oxygen-containing functional groups played a crucial role in optimizing active sites and tuning properties. Density functional theory calculations indicated that both the surfaces of $NaNbO_3$ and $CeO_2$ exhibit similarly low theoretical overpotentials for this reaction, with $CeO_2$ further enhancing the catalyst by facilitating electron transfer. These findings underscore the potential of metallic oxide heterostructures modifying C-based electrocatalysts as promising materials for *in situ* $H_2O_2$ electrogeneration, offering valuable insights for future applications and advancements in this field.

Moura *et al.* (Moura et al., 2023) investigated the influence of different crystalline phases of the same oxide ($MnO_2$) on the catalytic activity and showed that the crystal structure and morphology of the modified $MnO_2$/C yielded distinct performance for the 2-electron ORR. Nosan *et al*. [104] reported different nickel-functionalities on reduced graphene oxide (rGO) surfaces. The catalysts were prepared at various heat treatment temperatures in a slightly oxidizing atmosphere. The results revealed alterations in the nickel/oxygen functionalities, resulting in different electrochemical performance, stability, and selectivity for $H_2O_2$ production. The NiO-rich catalyst – (Ni@rGO treated at 900ºC) reached the highest % $H_2O_2$ production with 89% efficiency.

Coupling metal nanoparticles with metallic oxides is another interesting strategy to enhance the $H_2O_2$ electrogeneration on carbon materials (Fortunato et al., 2022; Kronka et al., 2023). Kronka *et al.* (Kronka et al., 2023) have achieved a highly efficient catalyst for $H_2O_2$ production by anchoring gold nanoparticles (Au NPs) onto a hybrid substrate composed of $ZrO_2$ and Printex L6 carbon (Au-$ZrO_2$/PL6C). The Au-$ZrO_2$/PL6C catalyst exhibited a remarkable 97% selectivity towards $H_2O_2$ electrogeneration. It also demonstrated improved activity in terms of ORR onset potential to more positive values compared to Au/PL6C (selectivity of 80%). These enhanced catalytic properties were attributed to the synergistic effect between the gold nanoparticles and the $ZrO_2$/PL6C hybrid support. This synergism facilitated efficient electron transfer and provided a favorable environment for the $H_2O_2$ generation during the ORR.

Summarily, modifying carbon with metallic oxides can enhance the catalytic activity for hydrogen peroxide electrogeneration from the oxygen reduction reaction by providing new active sites, improving conductivity and hydrophilicity, as well as regulating active sites. These improvements result in higher electrocatalytic efficiency and this is a promising approach for electrochemical applications involving $H_2O_2$.

In general, the literature has been actively exploring methods to fine-tune ORR activity in carbon-based catalysts, striving to attain optimal outcomes with high selectivity for the ORR 2-electron mechanism. Despite these efforts, there are gaps in the fundamental understanding of the modifying effects and structural diversity of such carbon catalysts. However, certain studies have started to address and demystify these aspects, primarily employing a combination of theoretical simulation tools and experiments to comprehend the origin of the activity.

## 4. Theoretical simulations for electrogeneration of $H_2O_2$

Computational simulations are very useful tools to understand what contributes to a great catalytic performance and selectivity of a given material at an atomic level. The obtained data can either complement the experimental analysis or provide entirely new information that would not be accessible otherwise. In this sense, density functional theory (DFT) calculations are often performed to unravel catalytic trends, providing results in the form of adsorption and reaction energies, Gibbs free energies, charge transfer processes happening on the catalytic surfaces, electronic properties such as band structure and density of states, and structural changes that might happen over the course of a given reaction (Lucchetti et al., 2021).

An important milestone for studying the ORR with DFT calculations was the development of the computational hydrogen electrode (CHE) model, by Nørskov *et al.* (J. K. Nørskov et al., 2004). In this case, the experimental standard reduction potentials are employed so that theoretical overpotential values are closer to the experimentally observed onset potential, providing a straightforward comparison with the experiments. Besides, only the reaction intermediate steps need to be considered, reducing computational costs. Other electrochemical cell phenomena can be described with this model in a simplified way, such as pH and applied potential effects. This model revolutionized the field and has been extensively discussed and used in the literature ever since.

The Gibbs free energy of reaction intermediates has become particularly useful as a catalytic descriptor (J. K. Nørskov et al., 2004), and since then, other descriptors have been proposed in the literature as well (Li et al., 2020; Lucchetti et al., 2021; Chaudhary et al., 2023). With unified descriptors calculated systematically, the analysis of different materials now became possible on a large scale and with comparable results, which has always been a major obstacle for meaningful correlations among different computational studies (Lucchetti et al., 2021). This effort from the computational chemistry scientific community to work under the same framework in a systematic way accounts for the progress achieved in the last decade - thousands of different materials have been thoroughly investigated with DFT alongside experimental validation.

For the last decade, the 4-electron ORR pathway was a major focus of most studies using DFT, as shown in a review from 2021 (Lucchetti et al., 2021). The $H_2O_2$ generation was usually taken as an unwanted parallel process to be avoided in fuel cells, and therefore, not often addressed in the literature. Some pioneering works in that direction have been carried out by Siahrostami *et al.* (Siahrostami et al., 2013), Chen *et al.* (Chen et al., 2018),

Lu *et al.* (Lu et al., 2018), and Koh *et al.* (Koh et al., 2023) focused on the 2-electron ORR for $H_2O_2$ electrogeneration, using DFT calculations to show the role of specific types of defects and oxygenated functions for the catalytic activity of carbon-based materials, as well as of the coordination environment in single-atom catalysts. In fact, oxygenated functions or borders seem to be the main catalytic sites to promote this reaction across different carbon-based materials, including nanotubes (Chu et al., 2023). After that, there has been a shift of focus towards the $H_2O_2$ electrogeneration in this field, as can be seen by several very recent works in the literature (Mavrikis et al., 2021; Byeon et al., 2023; Gao et al., 2023; Zhang et al., 2023b).

As described in section 3.1, doping carbon materials with nitrogen, or a combination of nitrogen and different atoms, especially metals, has been a good strategy to improve the catalytic activity for the $H_2O_2$ generation. Carbon nanotubes, following that same trend, can also be improved with nitrogen doping, as shown by a combined experimental and theoretical work from Hu *et al.* (Hu et al., 2023b). Oxygenated groups (such as -COOH) can also be combined with nitrogen functionalization with a positive effect of performance enhancement for this reaction in carbon-based materials, as observed with DFT calculations (Sun et al., 2023a). Recently, Li *et al.* (Li et al., 2021b) synthesized a B, N co-doped carbon aerogel with remarkable performance for $H_2O_2$ electrogeneration, where the DFT simulations showed that hBN catalytic sites coupled with neighboring pyridinic-N synergistically were on top of the activity volcano plot, due to lowering the energy barrier for *OOH formation and the B 2p electrons strongly interacting with the O 2s electrons. The mechanism selectivity can also be fine-tuned and modulated by the action of the selected functionalization combined with nitrogen doping, as shown by Cheng *et al.* (Cheng et al., 2023). In this work, the ORR activity could be switched in N-doped hollow mesoporous carbon spheres by either the addition of P (increasing the 2-electron selectivity) or S atoms (increasing the 4-electron selectivity).

Recently, computational simulations have shed some light on which specific nitrogen sites might contribute to the catalytic selectivity. For instance, in an experimental and theoretical work from Peng *et al*. (Peng et al., 2023), the authors noted that the coordination of pyrrolic-N sites promotes a favorable adsorption of the *OOH intermediate species, therefore accounting for the outstanding performance observed. Dual pyrrolic-N sites, specifically, were able to promote the reaction with a remarkably low overpotential ($\Delta G_{*OOH}$

= 0.05 eV). Besides DFT calculations, the authors also performed *ab initio* molecular dynamics AIMD simulations to investigate the effects of the alkaline aqueous medium, showing that the higher reaction barriers for the 4-electron mechanism hinder this route and promote the observed $H_2O_2$ selectivity.

M-N (metal-nitrogen) or even O/N-M (metal, oxygen and/or nitrogen) moieties have also been extensively explored in the literature (Lucchetti et al., 2021), and particularly promising results have been recently obtained with moieties based on Ni (Zhang and Xiao, 2020; Chaudhary et al., 2023; Yue et al., 2023), adjacent co-doping Ni-Fe (Zhang et al., 2023c), Co (Yan et al., 2023), and Mo (Dong et al., 2023), with high efficiency and selectivity for the 2-electron ORR. N-doped carbon nanotube-encapsulated nickel nanocatalysts were also able to promote the direct ORR to hydroxyl radical and this material was employed for BPA degradation with ~100% efficiency, circumventing the Fenton process intrinsic limitations (Li et al., 2023).

Graphitic carbon nitride also seems a promising new carbon-based material for $H_2O_2$ electrogeneration (Teng et al., 2021). In a recent work (Feng et al., 2023), g-$C_3N_4$ was further enhanced with terminal −CN groups, and the DFT calculations indicated that this tertiary N site was an electron-rich region. This structure was not only very selective for the 2-electron ORR, but it was also capable of promoting its conversion to hydroxyl radical, which is highly desirable for water treatment applications. In a similar direction, Ren *et al.* (Ren et al., 2023b) synthesized atomically dispersed Mn on aryl amino substituted g-$C_3N_4$, a modification that enabled the direct $H_2O_2$ generation via water oxidation reaction (WOR) from seawater, under visible light irradiation.

This scenario is very relevant within the context of pollutant degradation since most emerging contaminants end up in the ocean as well (He et al., 2023). With the aid of computational simulations, the authors were able to identify the molecular orbitals involved in the process, as well as the HOMO-LUMO dominant transition (474 nm) on these materials, in very good agreement with the experimental results where the reaction was promoted with a LED lamp with a 427 nm wavelength. The authors emphasize that the computational analysis indicated that photoexcited electrons and corresponding holes are actually localized in the vicinity of the aryl amino group, demonstrating the importance of also considering the metal neighbouring sites for ORR and WOR mechanisms. Yang *et al.* (Yang et al., 2023a) synthesized ultra-low content Pt single atoms on g-$C_3N_4$ nanosheets and performed GGA-PBE calculations modeling the g-$C_3N_4$ porous structure with Grimme's D3 vdW correction. Unlike pure g-$C_3N_4$, the DFT calculations showed that when a Pt atom is adsorbed on the porous structure and bonded to three nitrogen atoms, this single-atom catalytic site promotes the *OOH adsorption and, consequently, the $H_2O_2$ formation with zero overpotential. This is a remarkable result, given that most catalysts still have a low overpotential value associated with a potential determining reaction step. It also shows how the coordination environment can be a determinant factor for the preferred pathway since Pt-doped carbon materials are usually selective towards the 4-electron ORR.

It has been suggested in the literature (Anantharaj et al., 2021) that isolating ORR active sites by surrounding those with inert sites is a good strategy to preserve O-O bond and promote the 2-electron mechanism. This has also been corroborated by a theoretical work from Wei *et al.* (Wei et al., 2022b), where the authors observed that single Pd sites ($PdC_4$) are very active and selective for this mechanism, while the catalytic activity diminishes as Pd atoms clusterize over defects on the carbon network ($C_4Pd_x$, x = 2, 3, 4, 5). This shows why it is important to take the coordination environment into account. Another example in that direction is a work carried out by Wang *et al.* (Wang et al., 2020a) where DFT calculations revealed that, even though Fe-doped carbon materials are usually selective towards the 4-electron mechanism, the presence of OH axial ligands on the metal center shift the selectivity of the catalysts towards the $H_2O_2$ generation. These theoretical studies show that fine-tuning the catalytic sites at an atomic level might be a promising direction for the optimization of novel materials.

The functionalization of carbon-nitrogen materials with other organic structures to form electron donor-acceptor pairs is another interesting strategy (Sha et al., 2023), as is the combination of carbon-based and carbon-nitrogen structures as a way to fine-tune its molecular structure and consequently, the catalytic selectivity towards the reaction of interest (Luo et al., 2023). In these cases, DFT simulations can provide valuable insights into the charge-transfer properties of the obtained interfaces (Luo et al., 2023).

Among other carbon-based materials that have been explored recently with computational methods for the $H_2O_2$ electrogeneration, it is also worth mentioning the use of conjugated microporous polymers (Yang et al., 2023c), covalent organic frameworks (Huang et al.; Yang et al., 2023b; Zhou et al., 2023), metal-organic frameworks (Zhang et al., 2023a) and well-defined molecular metal phthalocyanines (Fan et al., 2023b), all with outstanding performance, and eventually with previous theoretical screening employed as a way to guide and optimize the experimental synthesis (Fan et al., 2023b). The advantage of this kind of carbon-based material is that there are numerous possibilities of elementary building blocks that can be specifically designed in order to regulate the physical and electronic structure to achieve high-performance catalysts (Ling et al., 2023).

Finally, other advanced carbon-based materials can still be explored for this application. A recent example is a boron-doped defective nanocarbon electrocatalyst synthesized from fullerene (C60) and boric oxide as the precursors from Shen *et al.* (Shen et al., 2023). The combination of topological pentagon defects and boron dopants yielded catalytic sites on top of the volcano activity top, with close-to-ideal $\Delta G_{*OOH}$ values and with a $H_2O_2$ production rate of 247 mg $L^{-1}$ $h^{-1}$. Fullerenes have already been extensively studied in the theoretical literature, and this work shows the new possibilities for using computational methods to investigate fullerene-based catalysts for the 2-electron ORR.

A few challenges still remain for accessing electrochemical processes with computation methods, such as the coordination environment, the complexities of solvent, ions in solution, and a reliable and cheap pH description, to mention a few (Di Liberto et al., 2023; Exner, 2023). For instance, Ramaswamy and Mukerjee (Ramaswamy and Mukerjee, 2011) proposed that, in alkaline media, an outer-sphere electron transfer mechanism is what might cause the 2-electron ORR to happen, with the change in pH affecting the electric double-layer structure entirely and consequently shifting the preferred reaction pathway. This type of contribution can be the determining factor for the mechanism selectivity, and it cannot be addressed in any way within the CHE framework.

Only recently a model has been proposed to access this kind of effect with DFT calculations by Kelly *et al.* (Kelly et al., 2020) and Li *et al.* (Li et al., 2021a), describing the local changes in the double-layer with a saw-tooth explicitly applied potential, keeping the surface atoms froze and allowing the reaction intermediates to relax. This model can indeed describe this kind of effect in very good agreement with experimental observations where only the CHE and its mathematical pH correction would fail, but the calculations are considerably more expensive. Zhang *et al.* (Zhang et al., 2022b) also were able to investigate the behavior of the electric double-layer with different pH values, using AIMD simulations. The authors set up the medium pH by adding the equivalent number of cations/protons to account for a $V_{\mathrm{RHE}}$ (potential relative to RHE) = −1 *V*. They were able to observe that, under these conditions, $Na^+$ cations on the double-layer have a "protective" effect, keeping protons away from the $H_2O_2$ molecule during the ORR and preventing its dissociation. Another work in that direction also highlighted the effect of $Cl^-$ ions in a P-doped carbon catalyst, stabilizing the adsorbed $O_2$ molecule and the *OOH intermediate and therefore promoting the 2-electron ORR (He et al., 2023). These recent studies show that continuous development and methodology refining are required so that computational simulations can approach the actual electrochemical cells' environment, as well as correctly describe the outliers or exceptions from the previously established methods.

After the major breakthroughs achieved with the CHE framework - high-throughput screening of catalysts, thousands of calculations being performed within a unified method and with replicable results, and even the development of new catalysts entirely from computational calculations (Chen et al., 2023a) - we are currently under a new paradigm in this field. There is already a lot of effort and resources being employed by the scientific community worldwide to make this data available and to speed up the development of novel and better catalysts. Some noteworthy examples of that are the Open Catalysis Project (https://opencatalystproject.org/), the Catalysis Hub (https://www.catalysis-hub.org/), and the Materials Project Catalysis Explorer (https://next-gen.materialsproject.org/catalysis). With the ever-growing amount of data being generated in this field, we are now moving towards the utilization of machine learning and artificial intelligence (ML/IA) methods to 1) try to identify the same catalytic trends already established with DFT with computationally cheaper algorithms, and 2) train new models that are capable of extrapolating from the available data and predict entirely new structures of groundbreaking catalysts.

Regarding the $H_2O_2$ electrogeneration, some recent works in the literature have employed ML to screen MN (metal-nitrogen) functionalized carbon materials (Chen et al., 2023b). For instance, Wang *et al.* (Wang et al., 2023c) used ML to investigate the $O_2$ adsorption on Ni(II), Co(II), Cu(II), Fe(II), Fe(III), and Mn(II) single-atom catalysts supported on 15 different N−C substrates and proposed a new descriptor derived by independence screening and sparsifying operator (SISSO) (Ouyang et al., 2018), an approach that can be used for small training sets. Zhang *et al.* (Zhang et al., 2023d) screened 690 different single-atom catalysts for the 2-electron ORR with different ML algorithms and identified 4 new promising materials, namely Zn@Pc-$N_3C_1$, Au@Pd-$N_4$, Au@Pd-$N_1C_3$, and Au@Py-$N_3C_1$, with remarkably low overpotentials. The authors also note that the catalytic activity prediction of these materials using ML is about 500 thousand times faster than that based on DFT.

Machine learning seems to be a new and promising approach to finding new catalysts (Chen et al., 2023c) but the best and most important features (among element identity, number of electrons in d or p orbitals, Bader charge, d-band center, bond lengths, enthalpy of formation, energy above Hull, adsorption energies and structural parameters, just to list some of the possibilities) and models to be chosen are still being studied and developed (Tamtaji et al., 2022; Chen et al., 2023c). Another argument is whether the best descriptors should be correlated with actual physical interpretations or if one should rely on the "black box" nature of some ML methods (Tamtaji et al., 2022). Finally, a major obstacle is the amount of data required to train those models - not only a high-quality and very large database is required, but specifically negative examples must also be present to train most of the ML algorithms so that the results can be reliable. Unfortunately, "negative results" are highly swept under the rug in the scientific literature as it is now, and only the best and most promising results are published and made available. Theoretical calculations are extremely valuable and it is evident how far we have come with these tools, especially over the last two decades. The future of this field is also being shaped by that, and what we want and need for the development and discovery of catalytic materials will depend on how well the scientific community is able to improve computational methods and converge to a common ground from now on.

## 5. Environmental applications

### *5.1. Treatment of effluents through advanced oxidative electrochemical processes*

#### 5.1.1. Electro-Fenton, photo-electro-Fenton, and solar-electro-Fenton process

Electrochemical advanced oxidation processes (EAOPs) have already demonstrated great potential for organic pollutants degradation in aqueous medium. The $H_2O_2$ molecule has a crucial role in this process and it is considered a green reagent since during the degradation process $H_2O_2$ breaks down into water and oxygen. However, there are some limitations in this kind of process, for example, certain pollutants can be partially degraded into other harmful organic structures or nitro compounds, cyanides, and some sulfate groups. Furthermore, some organic molecules are more recalcitrant, which means that an additional step is often required to completely oxidize them. In this context, Fenton reactions can be a good alternative, since the highly reactive radical •OH is generated. •OH radical is a powerful oxidizing agent, capable of attacking and breaking down an extensive range of organic molecules (Brillas, 2012; Clematis and Panizza, 2021; Gamarra-Güere et al., 2022).

Different kinds of electrochemical oxidative processes have been developed, namely:

1- Electro-Fenton: In this process, the Fenton reaction is coupled with electrochemical techniques and the $H_2O_2$ electrogeneration occurs *in situ*, along with Fe reduction.
2- Photo-electro-Fenton: in this case, light is used to activate the Fenton reaction, and to amplify the •OH production (Borràs et al., 2013).
3- Solar-electro-Fenton: similar to the previous item, but in this case solar photoactivation is employed (Skoumal et al., 2009).

Electro-Fenton (EF) processes are based on the $H_2O_2$ electrogeneration combined with iron ($Fe^{2+}$) oxidation under acid conditions, producing highly reactive •OH species. The reaction can be written as the equation below:

$$Fe^{2+} + H_2O_2 \rightarrow Fe^{3+} + \bullet OH + OH^- \quad (7)$$

The $Fe^{2+}$ species act as catalysts, being oxidized to ferric ions ($Fe^{3+}$) while generating •OH and hydroxide ions ($OH^-$) as products. In this reaction, a small quantity of $Fe^{2+}$ is added to the solution containing $H_2O_2$ to promote the Fenton reaction. The generated •OH is highly reactive and can attack and oxidize organic pollutants in the solution, breaking them down into smaller structures and less harmful compounds. The optimum pH value for this kind of process is around 3. This low pH is necessary for the efficient hydroxyl radicals generation through the Fenton process. Photo-electro-Fenton (PEF) and solar-electro-Fenton (SEF) are similar, except that UV irradiation increases the hydroxyl radical generation combining the electrochemical reactions with photochemistry principles (Oturan et al., 2021; Peleyeju and Viljoen, 2021). In both cases, additional hydroxyl radicals are formed by the action of ultraviolet (UV) irradiation, according to Eq. (8):

$$H_2O + hv\ (UV)\ \rightarrow \bullet OH + \bullet H \tag{8}$$

Since the UV irradiation enhances the production of hydroxyl radicals, its use leads to more efficient degradation processes (Brillas, 2013; Moreira et al., 2013). The SEF is a variation of the EF process where sunlight is the energy source for that. The reactions for this process are presented in Eq. (9), and (10) below:

$$Fe^{2+} + H_2O_2 \rightarrow \bullet OH + OH^- + Fe^{3+} \tag{9}$$

$$hv\ (UV/sunlight) + H_2O \rightarrow \bullet OH + \bullet H \tag{10}$$

It is evident that SEF offers a more renewable and environmentally friendly approach to the degradation of pollutants in comparison to using artificial UV light. Regardless of the method, PEF or SEF, however, UV irradiation has a decisive role in amplifying the •OH production (Brillas et al., 2009). Finally, another advantage of the EAOPs is that $Fe^{2+}$ species can be renewed on the cathode by the reduction reaction of $Fe^{3+}$, according to the following reaction described in Eq.(11) [176].

$$Fe^{3+} + e^- \leftrightarrow Fe^{2+} \tag{11}$$

As we continue to deal with a myriad of complex and persistent pollutants, the flexibility, efficiency, and environmental advantages of EAOPs make them an increasingly vital tool for wastewater treatment applications. One of the most compelling features of EAOPs is their efficiency in comparison with other traditional methods. They are particularly useful for degrading complex and highly stable organic molecules that are otherwise resistant to conventional treatment processes, and the combination with UV irradiation further enhances the processes at the same time it decreases their environmental footprint.

### 5.1.2 The radicals formed in each process

During the application of EAOPs, various radical species can be formed such as •OH, superoxide radicals (•$O_2$-), and other reactive oxygen species (ROS) (Wang et al., 2018). In EF, •OH and hydroperoxyl radicals ($HO_2$•) are the primary species generated during the process. These radicals are powerful oxidants, hydroxyl radicals being the most reactive and can quickly react with organic pollutants. These species are formed according to Eq. (12) and (13):

$$\bullet\, O_2 + H^+ + e^- \rightarrow \bullet\, O_2^- \quad (12)$$

$$HO_2^- + H^+ + e^- \rightarrow HO_2^\bullet \quad (13)$$

Additionally, the anode material employed in EF processes can also influence the types and amounts of radicals generated. Different anode materials can promote specific electrochemical reactions that result in the generation of distinct radicals. The active anodes (mixed metal oxide anodes – MOMs) are designed to generate ROS directly, including •OH. Depending on the medium, these anodes can also generate chlorine and sulfate radical species (Oliveira et al., 2018; Pointer Malpass and de Jesus Motheo, 2021). Non-active electrodes (DDB), on the other hand, do not directly generate hydroxyl radicals on their surface during the electrochemical process. Instead, they facilitate the electron transfer between the solution and electrode surface and the formation of chlorine species, if used in electrochlorination. Both active and non-active anodes have distinct advantages and applications in electrochemical processes (Colina-Márquez et al., 2009; Shin et al., 2019; de Mello et al., 2021; Santacruz et al., 2022; Soares et al., 2023).

Regarding the generation of other species like Cl• (chlorine radical) and $SO_4$• (sulfate radical), their presence and contribution to the organic pollutants degradation are highly dependent on specific reaction conditions. Chlorine radical (Cl•) generation may occur when chloride ions are present in an aqueous medium and undergo different reactions at the anode surface. This can happen in some electrochemical processes when chloride ions are naturally present in the water or when chlorine-based compounds are intentionally added to aid in disinfection (Wang and Wang, 2020; Dionisio et al., 2022). Likewise, sulfate radical ($SO_4$•) generation can happen when sulfate ions are present in the water. These radicals are potent oxidants and can also contribute to the organic pollutants degradation (Liu et al., 2019).

Other reactive species can be formed when an alcoholic medium is used, such as methoxy radicals and ethoxy radicals. Methoxy radicals (•$OCH_3$), are produced when the hydroxyl radical •OH removes a hydrogen atom from a methoxy group (-$OCH_3$), present in the solution. The ethoxy radicals (•$OC_2H_5$) can be formed similarly. These radicals are also highly reactive and can contribute to the degradation process. Their formation can be influenced by various factors such as pH, temperature, and the presence of metal catalysts. Therefore, optimizing the process conditions is essential to maximize the generation and effectiveness of these radical species to promote pollutant degradation (de Mello et al., 2022a; de Mello et al., 2022b; Fiori et al., 2022; Patidar and Srivastava, 2022; Santacruz et al., 2023). In this sense, electron paramagnetic resonance (EPR) spectroscopy, is a powerful analytical technique used for the characterization and detection of free radicals and other paramagnetic species. This method allows the investigation of the nature, concentration, and reactivity of free radicals (Santacruz et al., 2022; Webster, 2023).

### *5.2. Degradation of organic pollutants*

The production of organic wastewater across various sectors, including print and textile dyeing, agriculture, and healthcare, has become a major environmental issue. This kind of wastewater might contain a plethora of toxic organic pollutants, including dyes, pharmaceuticals, and pesticides, all of which possess high molecular bond energy and pose considerable challenges in terms of degradation (Lai et al., 2023). Consequently, the rising levels of environmental contamination and associated biological problems attributable to organic pollutants call for the development of environmentally responsible and sustainable methodologies to efficiently address organic wastewater treatment.

Pesticides, particularly, pose a significant environmental challenge. These compounds are extensively employed in agriculture to enhance crop yields, however, inadequate application practices and subsequent runoff may result in water pollution. Persistent pesticides can accumulate in aquatic ecosystems, affecting aquatic life and potentially entering the food chain (Syafrudin et al., 2021). Moreover, these chemicals may exhibit toxicity to non-target species, disrupt ecosystems, and even lead to the development of pesticide-resistant organisms. Addressing this issue requires comprehensive monitoring, improved agricultural practices, and the development of eco-friendly alternatives to minimize pesticide pollution and safeguard water quality (Aidoo et al., 2023; Brovini et al., 2023).

Dyes can also be very harmful to water bodies, affecting their aesthetic quality, elevating biochemical and chemical oxygen demand, inhibiting photosynthesis, and compromising plant growth. Moreover, the recalcitrant nature and bioaccumulative properties of these dye compounds may pose risks of toxicity, mutagenicity, and even carcinogenicity (Al-Tohamy et al., 2022; Ravishankar Patil et al., 2022).

Pharmaceuticals in the environment have the potential to exert toxicity at various levels within biological systems and their impacts on organisms remain incompletely elucidated. Certain pharmaceuticals can induce enduring and irreversible genetic alterations in microorganisms, even at low concentrations, leading to increased resistance. The combination of these pollutants into 'drug cocktails' often results in unpredictable toxicity. Additionally, specific pharmaceuticals have been categorized as endocrine-disrupting compounds (EDCs) due to their harmful impacts on the human endocrine system (Sires and Brillas, 2012; Martínez-Sánchez et al., 2022; Ortúzar et al., 2022).

The preceding observations emphasize the importance of completely removing these pollutants and their metabolic byproducts from aquatic ecosystems to prevent potential toxicity and alleviate other adverse health effects. The widespread occurrence of these pollutants calls for effective remediation strategies. To tackle the urgent challenge of removing and mitigating organic pollutants, various treatment approaches have been developed. The following section provides a comprehensive compilation of literature investigations focusing on the degradation of pesticides, dyes, and pharmaceuticals, using

Fenton-based processes with in situ electrogenerated $H_2O_2$, aiming to address the risks associated with toxic organic pollutants and explore sustainable solutions.

### 5.2.1. Wastewaters containing pesticides

Synthetic solutions polluted with pesticides like thiamethoxam (Meijide et al., 2016), pymetrozine (Fdez-Sanromán et al., 2020), parachlorophenol (Al-Zubaidi and Pak, 2020), lindane (Dominguez et al., 2018), a mixture of diuron and glyphosate (Rosa Barbosa et al., 2018), 2,4-D (García et al., 2014), and S-metolachlor (Guelfi et al., 2018) have been remediated with EF processes. Other pesticides such as atrazine (Komtchou et al., 2017) and carbaryl (Kronka et al., 2023) have been treated with the PEF process. A mixture of tebuthiuron and ametryn (Gozzi et al., 2017), bentazol (Guelfi et al., 2018), and asulam (Vigil-Castillo et al., 2023) were treated by a SPEF treatment (**Table 1**). All Fenton-based processes described in Table 1 were conducted with *in situ* electrogeneration of $H_2O_2$ through $O_2$ electroreduction. The $H_2O_2$ production is evidently influenced by various operational parameters, including the electrode configuration, applied current, and mass transport characteristics of the chemical species involved in both the chemical and electrochemical reactions inherent to the process (Martínez-Sánchez et al., 2022).

In this context, Komtchou *et al.* (Komtchou et al., 2017) observed that $H_2O_2$ electrogeneration is proportional to the applied current density. In their study, they employed graphitic carbon plates as the cathode and a BDD electrode as the anode within a reactor. They found that applying current densities of 2 mA $cm^{-2}$ and 18.2 mA $cm^{-2}$ led to $H_2O_2$ concentrations of 1.0 mg $L^{-1}$ and 4.1 mg $L^{-1}$, respectively. Interestingly, they observed that as the current density increased, so did the production of $H_2O_2$. However, it reached a maximum point beyond which no additional increase in $H_2O_2$ production was observed, likely due to its subsequent oxidation at the anode.

It has been noted that the electrogeneration of $H_2O_2$ is enhanced when materials with high surface area are employed, and other parameters such as excellent conductivity, large porosity, and high hydrophilicity also contribute positively (García et al., 2014; Kronka et al., 2023). For instance, Kronka *et al.* (Kronka et al., 2023), demonstrated that the incorporation of Au-$ZrO_2$ onto the surface of carbon Printex L6 resulted in a notable enhancement of $H_2O_2$ production. The improved selectivity of Au-$ZrO_2$/PL6C was ascribed to the synergistic interplay between anchored Au nanoparticles on the $ZrO_2$/PL6C hybrid support. This synergy facilitated electron transfer and promoted the metallic state of Au nanoparticles, ultimately boosting catalytic activity in $H_2O_2$ production.

**Table 1.** Literature review on pesticide degradation using Fenton-based processes with *in-situ* $H_2O_2$ electrogeneration from $O_2$ electroreduction.

| **Pesticides** | **Fenton-based Process** | **Electrodes** | | **Experimental parameters** | **Removal efficiency** | **Reference** |
|---|---|---|---|---|---|---|
| | | Cathode | Anode | | | |
| Atrazine $C_0$ = 100 µg $L^{-1}$ | Photoelectro-Fenton | Graphite | BDD | *j* = 2 mA $cm^{-2}$, 0.03 M $Na_2SO_4$, 0.1 mM $Fe^{2+}$, pH = 3.0, 23 ± 2 ºC, UV irradiation (λ = 254 nm). Working volume = 5.0 L. | 100% HPLC (90 min) 81.3% TOC (7 h) | (Komtchou et al., 2017) |
| Thiamethoxam $C_0$ = 60 mg $L^{-1}$ | Electro-Fenton | Carbon-felt | BDD | *j* = 14.3 mA $cm^{-2}$, 0.01 M $Na_2SO_4$, 0.1-0.3 mM $Fe^{2+}$, pH = 2.8. Working volume = 0.25 L. | 100% HPLC (10 min) 92% TOC (8 h) | (Meijide et al., 2016) |
| Organochlorine pesticides $C_0$ = 0.2 mg $L^{-1}$ | Electro-Fenton | Carbon-felt | BDD | *j* = 16.6 mA $cm^{-2}$, without addition of supporting electrolyte, 0.1 mM $Fe^{2+}$, pH = 7.0, 25 ºC. Working volume = 0.23 L. | 100% HPLC (4 h) 90% TOC (6 h) | (Dominguez et al., 2018) |
| Pymetrozine $C_0$ = 5 mg $L^{-1}$ | Electro-Fenton | Fe-doped carbon felt | BDD | *I* = 100 mA, 0.01 M $Na_2SO_4$, 0.3 mM $Fe^{2+}$, pH = 3.0. Working volume = 0.15 L. | 100% HPLC (2 h) 90% TOC | (Fdez-Sanromán et al., |
| | | | | | (5 h) | 2020 ) |
| Carbaryl<br>$C_0$ = 5-20 mg $L^{-1}$ | Photoelectro-Fenton | Au-$ZrO_2$/PL6C | BDD | $j$ = 50 mA $cm^{-2}$, 0.1 M $K_2SO_4$, 0.25 mM $Fe^{2+}$, pH = 2.5, 25 ºC, 5 W UVC lamp. Working volume = 0.15 L. | 90% TOC (1 h) | (Kronka et al., 2023 ) |
| Parachlorophenol<br>$C_0$ = 50 mg $L^{-1}$ | Electro-Fenton | Carbon-felt | Ti-$RuO_2$/$IrO_2$ | $j$ = 4.1 mA $cm^{-2}$, supporting electrolyte not reported, 0.59 mM $Fe^{2+}$, pH = 2.5, 25 ºC. Working volume = 1.0 L. | ~92% COD (2 h) | (Al-Zubaidi and Pak, 2020 ) |
| Lindane<br>$C_0$ = 5-10 mg $L^{-1}$ | Electro-Fenton | 3D carbon-felt | BDD | $j$ = 16.6 mA $cm^{-2}$, 0.05 M $Na_2SO_4$, 0.05 mM $Fe^{2+}$, pH = 3.0, 25 ºC. Working volume = 0.23 L. | 100% HPLC (15 min)<br>80% TOC (5 h) | (Dominguez et al., 2018 ) |
| Bentazon<br>$C_0$ = 0.208 mM | Solar photoelectro-Fenton | Carbon-PTFE | BDD, Pt, or Ti\|Ru$O_2$ | $j$ = 3.33-100 mA $cm^{-2}$, 0.05 M $Na_2SO_4$, 0.5 mM $Fe^{2+}$, pH = 3.0, 30 ºC, air flow rate = 1 L $min^{-1}$. Working volume = 0.13 L. | 100% HPLC (20 min)<br>89% TOC (4 h) | (Guelfi et al., 2019 ) |
| Tebuthiuron* + ametryn#<br>$C_0$ = 20 mg $L^{-1}$ (TOC) | Solar photoelectro-Fenton | Carbon-PTFE | BDD | $j$ = 25-100 mA $cm^{-2}$, 0.05 M $Na_2SO_4$, 0.5 mM $Fe^{2+}$, pH = 3.0, 35 ºC, liquid flow rate = 200 L $h^{-1}$. Working volume = 2.5 L. | 100% HPLC (240 min*, 120 min#)<br>53% TOC (6 h) | (Gozzi et al., 2017 ) |
| Diuron $C_0$ = 0.10 mM + glyphosate $C_0$ = 0.13 mM | Electro-Fenton | Carbon-felt | $Ti/Ru_{0.36}Ti_{0.64}O_2$ | $j$ = 0.45 mA cm$^{-2}$, 0.05 M $Na_2SO_4$, 1.0 mM $Fe^{2+}$, pH = 3.0, room temperature, air flow rate = 1 L min$^{-1}$. Working volume = 1.0 L. | 34% TOC (3 h) | (Rosa Barbosa et al., 2018) |
| 2,4-D $C_0$ = 276 mg L$^{-1}$ | Electro-Fenton | Carbon–PTFE | BDD | $I$ = 0.5-2.0 A, 0.05 M $Na_2SO_4$, 1.0 mM $Fe^{2+}$, pH = 3.0, 35 ºC, liquid flow rate = 200 L h$^{-1}$. Working volume = 2.5 L. | 83% TOC (7 h) | (García et al., 2014) |
| Asulam $C_0$ = 10 mg L$^{-1}$ | Solar photoelectro-Fenton | Carbon-felt | BDD | $I$ = 0.3 A, 0.11 mM $Na_2SO_4$, 0.3 mM $Fe^{2+}$, pH = 3.0, room temperature, air flow rate = 1 L min$^{-1}$. Working volume = 0.2 L. | 96% HPLC 90% TOC (2 h) | (Vigil-Castillo et al., 2023) |
| S-metolachlor $C_0$ = 79.5 mg L$^{-1}$ | Electro-Fenton | Carbon–PTFE | BDD | $j$ = 100 mA cm$^{-2}$, 0.05 M $Na_2SO_4$, 0.5 mM $Fe^{2+}$, pH = 3.0, 25 ºC, air flow rate = 1 L min$^{-1}$. Working volume = 0.1 L. | 100% HPLC (4 h) 95% TOC (9 h) | (Guelfi et al., 2018) |

### 5.2.2. Wastewaters containing dyes

Synthetic solutions contaminated with various dyes, including a mixture of tetracycline, methyl orange, congo red and methylene blue (O. Silva et al., 2022), methyl orange (Marquez et al., 2020), and methylene blue (Soto et al., 2020) have been effectively remediated using the EF process. On the other hand, dyes like ponceau SS (dos Santos et al., 2018), and reactive orange 16 (Cornejo et al., 2023), have been treated with the PEF process. Sunset yellow (Pinheiro et al., 2020) acid blue 29 (Fajardo et al., 2019), allura red AC (Thiam et al., 2015), and evans blue (Antonin et al., 2015a) were successfully treated using the SPEF treatment (see **Table 2**).

In photoassisted electro-Fenton processes, choosing the right UV source is paramount, as the degradation kinetics can be influenced by both the wavelength and intensity of the radiation. Thus, careful consideration of different types of irradiation sources is essential to optimize the process efficiency. For example, Paz *et al.* (Paz et al., 2020) conducted EF and PA-EF (simulated solar photoelectro-Fenton) experiments by immersing a pen-type mercury UV lamp with a wavelength of 254 nm in a 350 mL Orange II dye solution (91.23 mg $L^{-1}$). After 60 minutes, EF achieved an 88% discoloration, while PA-EF achieved 100%. However, it is crucial to consider the relationship between the selected lamp, the obtained results, and the costs of the overall process.

In this regard, SPEF represents an attractive and environmentally sustainable alternative compared to the costs associated with high-energy UV lamps. By harnessing natural sunlight, SPEF offers a more economically viable and eco-friendly option for driving the photoassisted electro-Fenton process, making it an appealing choice for wastewater treatment and pollutant removal.

**Table 2.** Key findings derived from the degradation of dye compounds via Fenton-based processes utilizing electrogenerated $H_2O_2$ from $O_2$ electroreduction.

| **Dye** | **Fenton-based Process** | **System** | | **Experimental parameters** | **Best results** | **Reference** |
|---|---|---|---|---|---|---|
| | | Cathode | Anode | | | |
| Orange II $C_0$ = 0.26 mM | Simulated solar photoelectro-Fenton | $WO_{2.72}$/Vulcan XC72 | Pt | $j$ = 100 mA $cm^{-2}$, 0.1 M $K_2SO_4$, 0.5 mM $Fe^{2+}$, pH = 3.0. Working volume = 0.35 L. | 100% HPLC (1 H) 91% TOC (6 h) | (Paz et al., 2020) |
| Methyl Orange Methylene Blue* Congo Red# Tetracycline& | Electro-Fenton and photocatalysis | Pt-black | $TiO_2$ nanotube photoanode | Undivided quartz cell – Irradiation: 4 W UV lamp; 0.1 M $Na_2SO_4$, 0.2 mM $Fe^{2+}$, pH = 1.5. Working volume = 0.25 L. | 97.34%, 95.36%*, 93.23%# and 73.80%& UV-Vis (1h) | (O. Silva et al., 2022) |
| $C_0$ = 20 mg $L^{-1}$ | | | | | | |
| Methyl Orange<br>$C_0$ = 20-50 mg $L^{-1}$(TOC) | Electro-Fenton and photoelectro-Fenton | 3D-like air-diffusion carbon | Ti\|Ir-Sn-Sb oxide anode | *j* = 10-25 mA $cm^{-2}$, 0.05 M $Na_2SO_4$, 0.5 mM $Fe^{2+}$, pH = 3.0. Working volume = 2.0 L. | 99% UV-Vis (30 min)<br>EF 66% TOC (7h)<br>PEF 94% TOC (7h) | (Marquez et al., 2020) |
| Methylene Blue<br>$C_0$ = 20 mg $L^{-1}$ | Electro-Fenton | Carbon-PTFE | DSA | *j* = 16.67 mA $cm^{-2}$, 0.1 M $Na_2SO_4$, 0.5 mM $Fe^{2+}$, pH = 3.0-6.0, 25 ºC, air flow rate = 150 mL $min^{-1}$. Working volume = 0.5 L. | 100% UV-Vis<br>(8 min) | (Soto et al., 2020) |
| Sunset Yellow<br>$C_0$ = 100 mg $L^{-1}$ | Solar photoelectro-Fenton | Carbon-PTFE | Pt | *I* = 100 mA, 0.13 M NaCl, 0.25 g of $Fe_3O_4$ as catalyst, pH = 3.0, 20 ºC, $O_2$ gas bubbled at 0.2 bar. Working volume = 0.35 L. | 100% UV-Vis (15 min) 71% TOC (90 min) | (Pinheiro et al., 2020) |
| Ponceau SS<br>$C_0$ = 0.19 mM | Photoelectro-Fenton | Carbon-PTFE | BDD | *j* = 100 mA $cm^{-2}$, 0.05 M $Na_2SO_4$, 0.5 mM $Fe^{2+}$, pH = 3.0, 35 ºC, air flow rate = 400 mL $min^{-1}$, liquid flow rate = 200 L $h^{-1}$. Irradiation: 160 W UVA lamp. Working volume = 2.5 L. | 98% TOC (6 h) | (dos Santos et al., 2018) |
| Reactive Orange 16<br>$C_0$ = 50 mg $L^{-1}$ | Photoelectro-Fenton | Graphite felt | Ti\|$RuO_2$ | *j* = 10 mA $cm^{-2}$, 0.05 M $Na_2SO_4$, 0.5 mM $Fe^{2+}$, pH = 3.0, air flow rate = 1 L $min^{-1}$. Irradiation: UVA lamp. Working volume = 2.0 L. | 37% TOC (6 h) | (Cornejo et al., 2023) |
| Acid Blue 29 | Solar photoelectro-Fenton | Carbon-PTFE | Ti\|$RuO_2$ | *j* = 33.3 mA $cm^{-2}$, 0.05 M $Na_2SO_4$, 0.5 mM $Fe^{2+}$, pH = 3.0, 35 ºC, liquid flow rate = | 100% HPLC (15 min) | (Fajardo et al., 2019) |
| $C_0$ = 118 mg $L^{-1}$ | | | | 300 mL $min^{-1}$. Working volume = 0.13 L. | ~100% COD (6 h) | |
| Allura Red AC $C_0$ = 115 mg $L^{-1}$ | Solar photoelectro-Fenton | Carbon-PTFE | Pt | *j* = 50 mA $cm^{-2}$, 0.05 M $Na_2SO_4$, 0.5 mM $Fe^{2+}$, pH = 3.0, 35 ºC, liquid flow rate = 200 mL $min^{-1}$. Working volume = 2.5 L. | 100% UV-Vis (40 min) 94% TOC (4 h) | (Thiam et al., 2015) |
| Evans Blue $C_0$ = 0.245 mM | Solar photoelectro-Fenton | Carbon-PTFE | Pt | *j* = 55.4 mA $cm^{-2}$, 0.05 M $Na_2SO_4$, 0.5 mM $Fe^{2+}$, pH = 3.0, 35 ºC, liquid flow rate = 300 mL $min^{-1}$. Working volume = 0.1 L. | 100% UV-Vis (20 min) 88% TOC (5 h) | (Antonin et al., 2015a) |
| Methyl Orange $C_0$ = $5x10^{-5}$ M | Photoelectro-Fenton | Grafite rod | $TiO_2/Ti_4O_7$ photo-anodes | 3 V, pencil-type UV lamp (λ= 365 nm) and saturated activated carbon, pH = 3.0, liquid flow rate = 110 mL $min^{-1}$, Working volume = 0.075 L. | 70% 60 min | (Becerril-Estrada et al., 2020) |
| Bright blue FCP $C_0$ =100 mg $L^{-1}$ | Electro-Fenton | Graphite cloth and Activated carbono as 3D type electrode | Carbon cloth | 0.5 A (corresponding to about 8 V) iron loaded ionic resin, pH 3, flow rate = 30 mL $min^{-1}$. Working volume = 0.25 L. | 100% 70 min UV-Vis | (Robles et al., 2020) |
| Orange II $C_0$ = 10 mg $L^{-1}$ | Electro-Fenton | half-cylin-ders of porous carbon | half-cylin-ders of porous carbon | 2.27 $mAcm^{-2}$ (corresponding to about 8 V) iron loaded ionic resin,pH 3, flow rate = 50 mL $min^{-1}$. Working volume = 2 L. | 83% UV-Vis 90 min | (García-Espinoza et al., 2021) |
| Methyl Orange $C_0$ = 20 mg $L^{-1}$ | Electro-Fenton | Modified carbon paste electrode | Pt wire | *V= -0.8 V vs Ag\|AgCl,.* Working volume = 0.02 L. | 99% UV-Vis 120 min | (Ortiz-Martínez et |
| | | | | | | al., 2021) |

### 5.2.3. Wastewaters containing pharmaceuticals

Various synthetic solutions contaminated with different pharmaceuticals, such as ibuprofen (Shi et al., 2021), ofloxacin (Xu et al., 2023a), sulfamerazine (Deng et al., 2020), and predinisolone (Mohammadi et al., 2023) have been effectively treated using the EF process. Conversely, pharmaceuticals like sulfamethazine (Wang et al., 2020e), paracetamol (Pinheiro et al., 2019), bronopol (Ye et al., 2019), and carbenicillin (Ren et al., 2023a) were oxidized through the PEF process. A mixture of sulfamethoxazole and trimethoprim (Murillo-Sierra et al., 2018), erythromycin (Pérez et al., 2017), ciprofloxacin (Antonin et al., 2015b), trimethoprim (Zhang et al., 2016), levofloxacin (Coria et al., 2018), and sulfanilamide (El-Ghenymy et al., 2013) were successfully treated using the SPEF process (see **Table 3**).

The selection of anode material significantly influences the degree of pollutant mineralization in aqueous media. Tables 1-3 demonstrate that $RuO_2$, $IrO_2$, $TiO_2$, graphite, and Pb have been extensively studied and demonstrated their effectiveness in removing pesticides, dyes, and pharmaceuticals. However, recent reports highlight BDD as the preferred anode material. Employing BDD for the removal of these organic pollutants has consistently yielded degradation ranging from 80 to 100%. This underscores the potential of BDD as a highly efficient option for achieving effective pollutant removal in water treatment applications.

In that sense, Ye *et al.* (Ye et al., 2019) examined the oxidation of bronopol through photoelectro-Fenton treatment, employing either a BDD or $RuO_2$-based anode. The combination of the BDD anode and a catalyzed cathode yielded the most promising results, achieving complete bronopol removal within 210 minutes and 94% mineralization after 360 minutes at 40 mA $cm^{-2}$. This exceptional performance was attributed to the synergistic action of $^{\bullet}OH$, BDD($^{\bullet}OH$), and sunlight, highlighting the potential of the BDD-based system for efficient pollutant degradation in water treatment applications. Ren *et al.* (Ren et al., 2023a) investigated the degradation of carbenicillin while comparing Pt and BDD as anode. BDD was selected due to its exceptional capacity to generate the largest amount of reactive M(•OH) according to Eq. (14):

$$M + H_2O \rightarrow M(OH) + H^+ + e^- \quad (14)$$

**Table 3.** Selected results obtained on the degradation of pharmaceuticals using Fenton-based processes with $H_2O_2$ electrogeneration in situ from $O_2$ electroreduction.

| **Pharmaceuticals** | **Fenton-based Process** | **Electrodes** | | **Experimental parameters** | **Removal efficiency** | **Reference** |
|---|---|---|---|---|---|---|
| | | Cathode | Anode | | | |
| Sulfamethoxazole + Trimethoprim<br>$C_0$ = 50 mg $L^{-1}$ of sulfamethoxazole + 11.1 mg $L^{-1}$ of trimethoprim | Solar photoelectro-Fenton | Carbon-PTFE | BDD | *j* = 33.3 mA $cm^{-2}$, supporting electrolyte: 0.05 M $Na_2SO_4$, 0.5 mM $Fe^{2+}$, pH = 3.0, 35 ºC, cathode fed with 1 L $min^{-1}$ of $O_2$. Working volume = 0.13 L. | 87.4%<br>TOC<br>(4 h) | (Murillo-Sierra et al., 2018) |
| Ibuprofen<br>$C_0$ = 10 mg $L^{-1}$ | Electro-Fenton | Hollow fiber-based structure incorporating carbon nanotubes | $IrO_2$ coating | *j* = 10 mA $cm^{-2}$, 0.05 mol $L^{-1}$ $Na_2SO_4$, 0.7 mM $Fe^{2+}$, pH = 7.0, flow rate of 0.6 L $min^{-1}$. Working volume = 0.2 L. | 94.8% HPLC (150 min)<br>41.8% TOC (120 min) | (Shi et al., 2021) |
| Sulfamethazine<br>$C_0$ = 0.2 Mm | Photoelectro-Fenton | Carbon black-polytetrafluoroethylene (PTFE) modified graphite felt | DSA with $IrO_2$ and $RuO_2$ | *j* = 100 mA $cm^{-2}$, 0.05 M $Na_2SO_4$, 0.3 mM $Fe^{2+}$, pH 3.0. Working volume = 0.2 L. | 83.5% TOC (4 h) | (Wang et al., 2020e) |
| Erythromycin $C_0$ = 0.225 mM | Solar photoelectro-Fenton | Graphite-felt | Platinized titanium plate | Cathodic current density: -0.16 mA $cm^{-2}$, 0.05 M $Na_2SO_4$, 0.5 mM $Fe^{2+}$, pH = 3.0, 35 ºC. Working volume = 1.4 L. | 69% TOC (5 h) | (Pérez et al., 2017) |
| Ciprofloxacin $C_0$ = 0.245 mM | Solar photoelectro-Fenton | Carbon-PTFE | BDD | $j$ = 33.3 mA $cm^{-2}$, 0.05 M $Na_2SO_4$, 0.5 mM $Fe^{2+}$, pH = 3.0, 25 ºC, $O_2$ flow rate = 300 mL $min^{-1}$. Working volume = 0.1 L. | 95% TOC (6 h) | (Antonin et al., 2015b) |
| Ofloxacin $C_0$ = 20 mg $L^{-1}$ | Electro-Fenton | Biogenic Fe-Mn oxides and carbon black (CB-Bio-$FeMnO_x$) | Ti/$PbO_2$ | $j$ = 50 mA $cm^{-2}$, 0.05 M $Na_2SO_4$, pH = 6.5. Working volume = 0.1 L. | 41% TOC (6 h) | (Xu et al., 2023a) |
| Sulfamerazine $C_0$ = 20 mg $L^{-1}$ | Electro-Fenton | Carbon felt | BDD sheet | $j$ = from 2.5 to 15 mA $cm^{-2}$, 0.05 M $Na_2SO_4$, 0.48 mM $Fe^{3+}$, pH = 3.0, $O_2$ flow rate = 300 mL $min^{-1}$. Working volume = 0.1 L. | 98.5% HPLC (90 min) 85.6% TOC (6 h) | (Deng et al., 2020) |
| Trimethoprim $C_0$ = 200 mg $L^{-1}$ | Solar photoelectro-Fenton | Activated carbon fiber | Ti\|$RuO_2$ | $j$ = 18 mA $cm^{-2}$, 0.05 M $Na_2SO_4$, 1.0 mM $Fe^{2+}$, pH = 3.0, 25 ºC, $O_2$ flow rate = 60 mL $min^{-1}$. Working volume = 0.125 L. | 80% TOC (6 h) | (Zhang et al., 2016) |
| Levofloxacin $C_0$ = 30 mg $L^{-1}$ | Solar photoelectro-Fenton | Carbon cloth\|graphite-felt | Pt | $E_{cat}$ = -0.30 V/SCE, 0.05 M $Na_2SO_4$, 0.5 mM $Fe^{2+}$, pH = 3.0, 25 ºC, liquid flow rate = 3 L $min^{-1}$. Working volume = 6.0 L. | 100% TOC (3 h) | (Coria et al., 2018) |
| Paracetamol $C_0$ = 157 mg $L^{-1}$ | Photoelectro-Fenton | 1% $CeO_2$ HARN/Vulcan XC-72 | BDD | $E_{cat}$ = -2.7 V/Ag/AgCl, 0.1 M $K_2SO_4$ + 0.1 M $H_2SO_4$, 1.0 mM $Fe^{2+}$, pH = 3.0, 35 | 98.4% TOC (6 h) | (Pinheiro et al., 2019) |
| | | | | ºC. Working volume = 0.35 L. | | |
| Bronopol $C_0$ = 0.28 mM | Photoelectro-Fenton | (Co, S, P)-modified MWCNTs | $RuO_2$ or BDD | $j$ = 40 mA cm$^{-2}$, 0.05 M $Na_2SO_4$, 0.5 mM $Fe^{2+}$, pH = 3.0, 35 ºC, liquid flow rate = 180 L h$^{-1}$. Working volume = 2.5 L. | 94% TOC (6 h) | (Ye et al., 2019) |
| Ibuprofen $C_0$ = 10 mg L$^{-1}$ | Electro-Fenton | Carbon-PTFE | $IrO_2$ sleeve | $j$ = 8 mA cm$^{-2}$, 0.05 M $Na_2SO_4$, 0.7 mM $Fe^{2+}$, pH = 3.0, air flow rate = 0.6 L min$^{-1}$. Working volume = 0.2 L. | ~50% TOC (2 h) | (Shi et al., 2022) |
| Sulfanilamide $C_0$ = 239 mg L$^{-1}$ | Solar photoelectro-Fenton | Carbon-PTFE | Pt | $j$ = 50-150 mA cm$^{-2}$, 0.05 M $Na_2SO_4$, 0.25-5.0 mM $Fe^{2+}$, pH = 3.0, 35 ºC, liquid flow rate = 200 L h$^{-1}$. Working volume = 2.5 L. | 92% TOC (5 h) | (El-Ghenymy et al., 2013) |
| Prednisolone $C_0$ = 20 mg L$^{-1}$ | Electro-Fenton | rGO/$SnO_2$/SSM | Pt | $I$ = 410 mA, 0.05 M $Na_2SO_4$, 1 g of rGO/$Fe_3O_4$/CF, pH = 4.5, room temperature. Working volume = 0.1 L. | 84% TOC (2 h) | (Mohammadi et al., 2023) |
| Carbenicillin $C_0$ = 0.049 mM | Photoelectro-Fenton | Nimesh\|C-PTFE | Pt or BDD | $I$ = 100 mA, 0.01 M $Na_2SO_4$, 0.5 mM $Fe^{2+}$, pH = 3.0, 25 ºC, air flow rate = 0.6 L min$^{-1}$. Working volume = 0.15 L. | 100% HPLC (13 min) 84% TOC (6 h) | (Ren et al., 2023a) |
| Triclosan $C_0$ = 10 mg L$^{-1}$ | Electro-Fenton | Carbon cloth | Carbon cloth | V= -0.8 V vs Ag\|AgCl, ionic resin as iron supporter, sequential circulation and polarization stage, 30 mL min$^{-1}$. Working volume = 0.2 L. | 100% HPLC (85 min) | (García-Espinoza et al., 2019) |

Thus, the EF processes have showcased efficiency in oxidizing organic contaminants in synthetic solutions. Significantly, the BDD anode surpasses Pt and DSA in oxidation capacity by generating a greater quantity of reactive radicals. Additionally, Carbon-PTFE-GDE cathodes demonstrate superior $H_2O_2$ production when compared to activated carbon fiber and carbon felt cathodes. However, a challenge emerges in the mineralization process, attributed to the gradual and complete photolysis of Fe(III)-carboxylate complexes. In this context, solar prepilot flow plant provide more practical insights compared to conventional tank reactors. For the Carbon-PTFE-GDE, the optimal $Fe^{2+}$ concentration was approximately 0.50 mM. Superior performance is consistently observed at higher contaminant concentrations and lower current densities, leading to a significant reduction in the rates of parasitic reactions. These reactions may envolve the dimerization reaction of •OH on the surface of the anode and/or in the reaction medium (Eq. (15)). Furthermore, there is the destruction reaction of •OH when reacting with $H_2O_2$, forming $HO_2^{\bullet}$ (Eq. (16)), and the reaction of •OH with $Fe^{2+}$ through reaction (Eq. (17)).

$$2^{\bullet}OH \rightarrow 2H_2O_2 \tag{15}$$

$$H_2O_2 + {}^{\bullet}OH \rightarrow HO_2^{\bullet} + H_2O \tag{16}$$

$$Fe^{2+} + {}^{\bullet}OH \rightarrow Fe^{3+} + OH^{-} \tag{17}$$

## 6. Conclusions and future outlook

Hydrogen peroxide is a powerful oxidizing agent, being used in hospital environments, paper and textile bleaching processes, chemical syntheses, and in water treatment plants based on advanced oxidative processes. The main methods of producing $H_2O_2$ are anthraquinone oxidation, direct synthesis of $H_2O_2$ using noble metals, and electrochemical methods. Currently, the most used is the anthraquinone oxidation process, but as it is a method that involves several steps and requires handling and storing $H_2O_2$, this method becomes expensive and not environmentally friendly. The direct synthesis of $H_2O_2$ uses noble metals, which makes this method more expensive. Furthermore, they pose serious risks of explosion, making this method unfeasible for large-scale production of $H_2O_2$. A promising, economically, and environmentally friendly alternative is the in situ production of $H_2O_2$ through the electrochemical method using the ORR. This methodology can be applied using carbon-based electrocatalysts, which are low cost, abundant, and high selectivity for $H_2O_2$.

Different carbon-based materials, such as graphene, graphite, carbon nanotube, and carbon black have already been studied and presented promising results for generating $H_2O_2$ by 2-electron ORR. However, as they are not ideal for this function, several modifications have already been made in order to optimize their performance. Carbon materials functionalized with metal atoms, for example, can be highly effective for 4-electron ORR, with the selection and synthesis methodology of the catalyst being fundamental in determining the selectivity of the ORR mechanism, tuning the catalyst surface with specific active sites for the 2-electron ORR to form hydrogen peroxide through electrochemical means. Another modification that stands out for 2-electron ORR is the addition of oxygen functional groups to carbon materials. This insertion can be carried out using different methodologies, which offer different advantages. The identification of catalytic active sites and modification of the carbon focus with active species can increase the efficiency and selectivity of $H_2O_2$ electrogeneration, paving the way for more sustainable and efficient electrochemical processes. Furthermore, the modification of carbon materials with metal oxides can increase the catalytic activity for the electrogeneration of $H_2O_2$ as it can provide new active sites, improving conductivity and hydrophilicity, as well as regulating the active sites. Therefore, it is clear that carbon materials are promising for the electrogeneration of $H_2O_2$ and different approaches can be used for this purpose.

To better understand the structure of different carbon materials, as well as the real effects of modifications used in these materials to improve their efficiency for generating $H_2O_2$ using the 2-electron ORR, many researchers have focused their efforts on carrying out theoretical simulations. Density functional theory calculations were carried out and provided results in the form of adsorption and reaction energies, Gibbs free energies, charge transfer processes occurring on the surfaces, among other information, which helped to understand how the modifications were able to favor the 2-electron ORR. Finally, different carbon-based electrocatalysts were applied to the degradation of pollutants using advanced oxidative electrochemical processes, being efficient in a wide class of pollutants (drugs, dyes, pesticides, and herbicides). The results presented in this review demonstrate how promising carbon materials are for electrogeneration of $H_2O_2$ using the 2-electron ORR. However, the practical implementation of this technology is still a challenge to be overcome. For this methodology to be used in large-scale productions, some obstacles must be overcome (Zhou et al., 2019a; Cordeiro-Junior et al., 2020; Salmeron et al., 2021; Wang et al., 2021b; Santos et al., 2022) such as:

1. Stability and durability: Carbon electrocatalysts can suffer from stability and durability problems during $H_2O_2$ electrogeneration processes. This compromises the performance and $H_2O_2$ production, generating additional costs. Thus, to apply these electrodes, studies with longer tests must be provided to determine the real stability/efficiency ratio of carbon electrocatalysts. Furthermore, the use of gaseous diffusion electrodes, which facilitate the transfer of electrons and $O_2$ mass to the cathode surface (Wang et al., 2021a), improving electrode efficiency, is highly recommended. However, the extensive use of this type of electrode can generate infiltrations and floods and affect its efficiency. Thus, longer stability tests and further investigation are even more necessary for this type of electrode;
2. Realistic tests: More realistic and robust parameters are required for both electrogeneration tests and application in effluent treatments, such as low electrolytes and high pollutant concentrations, as well as real matrices of reuse water and/or river waters;
3. Scalability and cost: For industrial applications, the scalability and cost-effectiveness of the electrogeneration process are crucial factors. Therefore, further studies should be aimed at designing reactors that promote high amounts of $H_2O_2$ production with low energy consumption.

**Acknowledgments**

The authors acknowledge financial support of the Brazilian research financing institutions: São Paulo Research Foundation (FAPESP - grants #2017/10118-0,

#2021/14394-7, #2021/05364-7, and #2022/10484-4), National Council for Scientific and Technological Development (CNPq - grant #303943/2021-1) and Coordenação de Aperfeiçoamento de Pessoal de Nível Superior (CAPES, 88887.354751/2019-00).

## References

Aidoo, O.F., Osei-Owusu, J., Chia, S.Y., Dofuor, A.K., Antwi-Agyakwa, A.K., Okyere, H., Gyan, M., Edusei, G., Ninsin, K.D., Duker, R.Q., Siddiqui, S.A., Borgemeister, C., 2023. Remediation of pesticide residues using ozone: A comprehensive overview. The Science of the total environment 894, 164933.

Al-Tohamy, R., Ali, S.S., Li, F., Okasha, K.M., Mahmoud, Y.A., Elsamahy, T., Jiao, H., Fu, Y., Sun, J., 2022. A critical review on the treatment of dye-containing wastewater: Ecotoxicological and health concerns of textile dyes and possible remediation approaches for environmental safety. Ecotoxicology and environmental safety 231, 113160.

Al-Zubaidi, D.K., Pak, K.S., 2020. Degradation of parachlorophenol in synthetic wastewater using Batch Electro-Fenton process. Materials Today: Proceedings 20, 414-419.

An, J., Feng, Y., Zhao, Q., Wang, X., Liu, J., Li, N., 2022. Electrosynthesis of H(2)O(2) through a two-electron oxygen reduction reaction by carbon based catalysts: From mechanism, catalyst design to electrode fabrication. Environmental science and ecotechnology 11, 100170.

Anantharaj, S., Pitchaimuthu, S., Noda, S., 2021. A review on recent developments in electrochemical hydrogen peroxide synthesis with a critical assessment of perspectives and strategies. Advances in Colloid and Interface Science 287, 102331.

Antonin, V.S., Garcia-Segura, S., Santos, M.C., Brillas, E., 2015a. Degradation of Evans Blue diazo dye by electrochemical processes based on Fenton's reaction chemistry. Journal of Electroanalytical Chemistry 747, 1-11.

Antonin, V.S., Lucchetti, L.E.B., Souza, F.M., Pinheiro, V.S., Moura, J.P.C., Trench, A.B., de Almeida, J.M., Autreto, P.A.S., Lanza, M.R.V., Santos, M.C., 2023. Sodium niobate microcubes decorated with ceria nanorods for hydrogen peroxide electrogeneration: An experimental and theoretical study. Journal of Alloys and Compounds 965, 171363.

Antonin, V.S., Parreira, L.S., Aveiro, L.R., Silva, F.L., Valim, R.B., Hammer, P., Lanza, M.R.V., Santos, M.C., 2017. W@Au Nanostructures Modifying Carbon as Materials for Hydrogen Peroxide Electrogeneration. Electrochimica Acta 231, 713-720.

Antonin, V.S., Santos, M.C., Garcia-Segura, S., Brillas, E., 2015b. Electrochemical incineration of the antibiotic ciprofloxacin in sulfate medium and synthetic urine matrix. Water research 83, 31-41.

Assumpção, M.H.M.T., De Souza, R.F.B., Rascio, D.C., Silva, J.C.M., Calegaro, M.L., Gaubeur, I., Paixão, T.R.L.C., Hammer, P., Lanza, M.R.V., Santos, M.C., 2011. A comparative study of the electrogeneration of hydrogen peroxide using Vulcan and Printex carbon supports. Carbon 49, 2842-2851.

Aveiro, L.R., da Silva, A.G.M., Antonin, V.S., Candido, E.G., Parreira, L.S., Geonmonond, R.S., de Freitas, I.C., Lanza, M.R.V., Camargo, P.H.C., Santos, M.C., 2018a. Carbon-supported MnO2 nanoflowers: Introducing oxygen vacancies for optimized volcano-type electrocatalytic activities towards H2O2 generation. Electrochimica Acta 268, 101-110.

Aveiro, L.R., Da Silva, A.G.M., Candido, E.G., Antonin, V.S., Parreira, L.S., Papai, R., Gaubeur, I., Silva, F.L., Lanza, M.R.V., Camargo, P.H.C., Santos, M.C., 2018b. Application and stability of cathodes with manganese dioxide nanoflowers supported on Vulcan by Fenton systems for the degradation of RB5 azo dye. Chemosphere 208, 131-138.
Barranco-Lopez, A., Moral-Rodriguez, A.I., Fajardo-Puerto, E., Elmouwahidi, A., Bailon-Garcia, E., 2023. Highly graphitic Fe-doped carbon xerogels as dual-functional electro-Fenton catalysts for the degradation of tetracycline in wastewater. Environmental research 228, 115757.
Becerril-Estrada, V., Robles, I., Martinez-Sanchez, C., Godinez, L.A., 2020. Study of TiO2/Ti4O7 photo-anodes inserted in an activated carbon packed bed cathode: Towards the development of 3D-type photo-electro-Fenton reactors for water treatment. Electrochim Acta 340, 135972.
Berl, E., 1939. A new cathodic process for the production of H2O2. . Trans. Electrochem. Soc. 76, 359-369.
Borràs, N., Arias, C., Oliver, R., Brillas, E., 2013. Anodic oxidation, electro-Fenton and photoelectro-Fenton degradation of cyanazine using a boron-doped diamond anode and an oxygen-diffusion cathode. Journal of Electroanalytical Chemistry 689, 158-167.
Brillas, E., 2012. Electro-Fenton, UVA Photoelectro-Fenton and Solar Photoelectro-Fenton Treatments of Organics in Waters Using a Boron-Doped Diamond Anode: A Review. J. Mex. Chem. Soc. 58, 239-255.
Brillas, E., 2013. A Review on the Degradation of Organic Pollutants in Waters by UV Photoelectro-Fenton and Solar Photoelectro-Fenton. Journal of the Brazilian Chemical Society.
Brillas, E., Sirés, I., Oturan, M., 2009. Electro-Fenton Process and Related Electrochemical Technologies Based on Fenton’s Reaction Chemistry. Chem. Rev. 109, Chem. Rev. 2009, 2109, 6570–6631.
Brovini, E.M., Moreira, F.D., Martucci, M.E.P., de Aquino, S.F., 2023. Water treatment technologies for removing priority pesticides. Journal of Water Process Engineering 53, 103730.
Bu, Y., Wang, Y., Han, G.-F., Zhao, Y., Ge, X., Li, F., Zhang, Z., Zhong, Q., Baek, J.-B., 2021. Carbon-Based Electrocatalysts for Efficient Hydrogen Peroxide Production. Adv. Mater. 33, 2103266.
Byeon, A., Choi, J.W., Lee, H.W., Yun, W.C., Zhang, W., Hwang, C.-K., Lee, S.Y., Han, S.S., Kim, J.M., Lee, J.W., 2023. CO2-derived edge-boron-doped hierarchical porous carbon catalysts for highly effective electrochemical H2O2 production. Applied Catalysis B: Environmental 329, 122557.
Campos-Martin, J.M., Blanco-Brieva, G., Fierro, J.L., 2006. Hydrogen peroxide synthesis: an outlook beyond the anthraquinone process. Angewandte Chemie 45, 6962-6984.
Chaudhary, P., Evazzade, I., Belosludov, R., Alexandrov, V., 2023. Computational Discovery of Active and Selective Metal-Nitrogen-Graphene Catalysts for Electrooxidation of Water to H2O2. ChemCatChem 15, e202300055.
Chen, K.-Y., Huang, Y.-X., Jin, R.-C., Huang, B.-C., 2023a. Single atom catalysts for use in the selective production of hydrogen peroxide via two-electron oxygen reduction reaction: Mechanism, activity, and structure optimization. Applied Catalysis B: Environmental 337, 122987.
Chen, S., Chen, Z., Siahrostami, S., Kim, T.R., Nordlund, D., Sokaras, D., Nowak, S., To, J.W.F., Higgins, D., Sinclair, R., Nørskov, J.K., Jaramillo, T.F., Bao, Z., 2018. Defective Carbon-Based Materials for the Electrochemical Synthesis of Hydrogen Peroxide. ACS Sustainable Chemistry & Engineering 6, 311-317.
Chen, Y., Cui, H., Jiang, Q., Bai, X., Shan, P., Jia, Z., Lu, S., Song, P., Feng, R., Kang, Q., Liang, Z., Yuan, H., 2023b. M-N4-Gr/MXene Heterojunction Nanosheets as Oxygen Reduction and Evolution Reaction Catalysts: Machine Learning and Density Functional Theory Insights. ACS Applied Nano Materials 6, 7694-7703.
Chen, Z., Qi, H., Wang, H., Yue, C., Liu, Y., Yang, Z., Pu, M., Lei, M., 2023c. The rational design of high-performance graphene-based single-atom electrocatalysts for the ORR using machine learning. Physical Chemistry Chemical Physics 25, 18983-18989.

Cheng, J., Lyu, C., Li, H., Wu, J., Hu, Y., Han, B., Wu, K., Hojamberdiev, M., Geng, D., 2023. Steering the oxygen reduction reaction pathways of N-carbon hollow spheres by heteroatom doping. Applied Catalysis B: Environmental 327, 122470.
Chu, L., Cang, L., Sun, Z., Wang, X., Fang, G., Gao, J., 2022. Efficient hydrogen peroxide electrosynthesis using anthraquinone covalently bonded CNT on superhydrophobic air breathing cathode. Journal of Cleaner Production 378, 134578.
Chu, L., Sun, Z., Cang, L., Wang, X., Fang, G., Gao, J., 2023. Identifying the roles of oxygen-containing functional groups in carbon materials for electrochemical synthesis of H2O2. Journal of Environmental Chemical Engineering 11, 109826.
Clematis, D., Panizza, M., 2021. Electro-Fenton, solar photoelectro-Fenton and UVA photoelectro-Fenton: Degradation of Erythrosine B dye solution. Chemosphere 270, 129480.
Colina-Márquez, J., Machuca-Martínez, F., Li Puma, G., 2009. Photocatalytic Mineralization of Commercial Herbicides in a Pilot-Scale Solar CPC Reactor: Photoreactor Modeling and Reaction Kinetics Constants Independent of Radiation Field. Environmental Science & Technology 43, 8953-8960.
Cordeiro-Junior, P.J.M., Kronka, M.S., Goulart, L.A., Veríssimo, N.C., Mascaro, L.H., Santos, M.C.d., Bertazzoli, R., Lanza, M.R.d.V., 2020. Catalysis of oxygen reduction reaction for H2O2 electrogeneration: The impact of different conductive carbon matrices and their physicochemical properties. Journal of Catalysis 392, 56-68.
Cordeiro-Junior, P.J.M., Martins, A.S., Pereira, G.B.S., Rocha, F.V., Rodrigo, M.A.R., Lanza, M.R.d.V., 2022. Bisphenol-S removal via photoelectro-fenton/H2O2 process using Co-porphyrin/Printex L6 gas diffusion electrode. Separation and Purification Technology 285, 120299.
Cordeiro Junior, P.J.M., Martins, A.S., Pereira, G.B.S., Rocha, F.V., Rodrigo, M.A.R., Lanza, M.R.d.V., 2022. High-performance gas-diffusion electrodes for H2O2 electrosynthesis. Electrochimica Acta 430, 141067.
Cordeiro Junior, P.J.M., Souto, R.S., Almeida, M.d.O., Pereira, G.B.S., Franco, M.A., Honorio, K.M., Rocha, F.V., Lanza, M.R.d.V., 2023. A combined approach toward enhancing 2-electron oxygen reduction through the incorporation of Pd-based complex into a carbonaceous matrix: Experimental and mechanistic-theoretical studies. Electrochimica Acta 460, 142543.
Coria, G., Perez, T., Sires, I., Brillas, E., Nava, J.L., 2018. Abatement of the antibiotic levofloxacin in a solar photoelectro-Fenton flow plant: Modeling the dissolved organic carbon concentration-time relationship. Chemosphere 198, 174-181.
Cornejo, O.M., Piña, F.J., Nava, J.L., 2023. Hybrid water treatment flow plant using hydrogen peroxide-based electro-activated persulfate and photoelectro-Fenton processes: The combustion of Reactive Orange 16 dye. Journal of Industrial and Engineering Chemistry 124, 558-569.
de Mello, R., Arias, A.N., Motheo, A.J., Lobato, J., Rodrigo, M.A., 2022a. Production of value-added substances from the electrochemical oxidation of volatile organic compounds in methanol medium. Chemical Engineering Journal 440, 135803.
de Mello, R., Motheo, A.J., Sáez, C., Rodrigo, M.A., 2022b. Combination of granular activated carbon adsorption and electrochemical oxidation processes in methanol medium for benzene removal. Electrochimica Acta 425, 140681.
de Mello, R., Rodrigo, M.A., Motheo, A.J., 2021. Electro-oxidation of tetracycline in methanol media on DSA®-Cl2. Chemosphere 273, 129696.
Deng, F., Li, S., Cao, Y., Fang, M.A., Qu, J., Chen, Z., Qiu, S., 2020. A dual-cathode pulsed current electro-Fenton system: Improvement for H2O2 accumulation and Fe3+ reduction. Journal of Power Sources 466, 228342.

Di Liberto, G., Barlocco, I., Giordano, L., Tosoni, S., Pacchioni, G., 2023. Single-atom electrocatalysis from first principles: Current status and open challenges. Current Opinion in Electrochemistry 40, 101343.
Dionisio, D., Rodrigo, M.A., Motheo, A.J., 2022. Electrochemical degradation of a methyl paraben and propylene glycol mixture: Interference effect of competitive oxidation and pH stability. Chemosphere 287, 132229.
Diouf, I., Dia, O., Diedhiou, M.B., Drogui, P., Toure, A.O., Lo, S.M., Rumeau, M., Mar/Diop, C.G., 2020. Electro-generation of hydrogen peroxide using a graphite cathode from exhausted batteries: study of influential parameters on electro-Fenton process. Environmental technology 41, 1434-1445.
Dominguez, C.M., Oturan, N., Romero, A., Santos, A., Oturan, M.A., 2018. Optimization of electro-Fenton process for effective degradation of organochlorine pesticide lindane. Catalysis Today 313, 196-202.
Dong, C., Wang, Z.-Q., Yang, C., Hu, X., Wang, P., Gong, X.-Q., Lin, L., Li, X.-y., 2023. Sequential electrocatalysis by single molybdenum atoms/clusters doped on carbon nanotubes for removing organic contaminants from wastewater. Applied Catalysis B: Environmental 338, 123060.
dos Santos, A.J., Martínez-Huitle, C.A., Sirés, I., Brillas, E., 2018. Use of Pt and Boron-Doped Diamond Anodes in the Electrochemical Advanced Oxidation of Ponceau SS Diazo Dye in Acidic Sulfate Medium. ChemElectroChem 5, 685-693.
Edwards, J.K., Pritchard, J., Lu, L., Piccinini, M., Shaw, G., Carley, A.F., Morgan, D.J., Kiely, C.J., Hutchings, G.J., 2014. The direct synthesis of hydrogen peroxide using platinum-promoted gold-palladium catalysts. Angewandte Chemie 53, 2381-2384.
El-Ghenymy, A., Cabot, P.L., Centellas, F., Garrido, J.A., Rodriguez, R.M., Arias, C., Brillas, E., 2013. Mineralization of sulfanilamide by electro-Fenton and solar photoelectro-Fenton in a pre-pilot plant with a Pt/air-diffusion cell. Chemosphere 91, 1324-1331.
Exner, K.S., 2023. Combining descriptor-based analyses and mean-field modeling of the electrochemical interface to comprehend trends of catalytic processes at the solid/liquid interface. Journal of Energy Chemistry 85, 288-290.
Fajardo, A.S., dos Santos, A.J., de Araújo Costa, E.C.T., da Silva, D.R., Martínez-Huitle, C.A., 2019. Effect of anodic materials on solar photoelectro-Fenton process using a diazo dye as a model contaminant. Chemosphere 225, 880-889.
Fan, M., Wang, Z., Sun, K., Wang, A., Zhao, Y., Yuan, Q., Wang, R., Raj, J., Wu, J., Jiang, J., Wang, L., 2023a. N□B□OH Site-Activated Graphene Quantum Dots for Boosting Electrochemical Hydrogen Peroxide Production. Adv Mater 35, e2209086.
Fan, M., Xu, J., Wang, Y., Yuan, Q., Zhao, Y., Wang, Z., Jiang, J., 2022. CO2 Laser-Induced Graphene with an Appropriate Oxygen Species as an Efficient Electrocatalyst for Hydrogen Peroxide Synthesis. Chemistry – A European Journal 28, e202201996.
Fan, W., Duan, Z., Liu, W., Mehmood, R., Qu, J., Cao, Y., Guo, X., Zhong, J., Zhang, F., 2023b. Rational design of heterogenized molecular phthalocyanine hybrid single-atom electrocatalyst towards two-electron oxygen reduction. Nature Communications 14, 1426.
Fdez-Sanromán, A., Acevedo-García, V., Pazos, M., Sanromán, M.Á., Rosales, E., 2020. Iron-doped cathodes for electro-Fenton implementation: Application for pymetrozine degradation. Electrochimica Acta 338, 135768.
Feng, S., Li, X., Kong, P., Gu, X., Wang, Y., Wang, N., Hailili, R., Zheng, Z., 2023. Regulation of the Tertiary N Site by Edge Activation with an Optimized Evolution Path of the Hydroxyl Radical for Photocatalytic Oxidation. ACS Catalysis 13, 8708-8719.
Feng, Y., Li, W., An, J., Zhao, Q., Wang, X., Liu, J., He, W., Li, N., 2021. Graphene family for hydrogen peroxide production in electrochemical system. The Science of the total environment 769, 144491.

Fiori, I., Santacruz, W., Dionisio, D., Motheo, A.J., 2022. Electro-oxidation of tetracycline in ethanol-water mixture using DSA-Cl(2) anode and stimulating/monitoring the formation of organic radicals. Chemosphere 308, 136487.
Fortunato, G.V., Bezerra, L.S., Cardoso, E.S.F., Kronka, M.S., Santos, A.J., Greco, A.S., Junior, J.L.R., Lanza, M.R.V., Maia, G., 2022. Using Palladium and Gold Palladium Nanoparticles Decorated with Molybdenum Oxide for Versatile Hydrogen Peroxide Electroproduction on Graphene Nanoribbons. ACS applied materials & interfaces 14, 6777-6793.
Fortunato, G.V., Kronka, M.S., dos Santos, A.J., Ledendecker, M., Lanza, M.R.V., 2020. Low Pd loadings onto Printex L6: Synthesis, characterization and performance towards H2O2 generation for electrochemical water treatment technologies. Chemosphere 259, 127523.
Gamarra-Güere, C.D., Dionisio, D., Santos, G.O.S., Vasconcelos Lanza, M.R., de Jesus Motheo, A., 2022. Application of Fenton, photo-Fenton and electro-Fenton processes for the methylparaben degradation: A comparative study. Journal of Environmental Chemical Engineering 10, 106992.
Gao, Y., Zhu, W., Wang, C., Zhao, X., Shu, M., Zhang, J., Bai, H., 2020. Enhancement of oxygen reduction on a newly fabricated cathode and its application in the electro-Fenton process. Electrochimica Acta 330, 135206.
Gao, Z., Zhu, Q., Cao, Y., Wang, C., Liu, L., Zhu, J., 2023. Design strategies of carbon-based single-atom catalysts for efficient electrochemical hydrogen peroxide production. Journal of Environmental Chemical Engineering 11, 109572.
García-Espinoza, J.D., Robles, I., Durán-Moreno, A., Godínez, L.A., 2021. Study of simultaneous electro-Fenton and adsorption processes in a reactor containing porous carbon electrodes and particulate activated carbon. Journal of Electroanalytical Chemistry 895, 115476.
García-Espinoza, J.D., Robles, I., Gil, V., Becerril-Bravo, E., Barrios, J.A., Godínez, L.A., 2019. Electrochemical degradation of triclosan in aqueous solution. A study of the performance of an electro-Fenton reactor. Journal of Environmental Chemical Engineering 7, 103228.
García, O., Isarain-Chávez, E., El-Ghenymy, A., Brillas, E., Peralta-Hernández, J.M., 2014. Degradation of 2,4-D herbicide in a recirculation flow plant with a Pt/air-diffusion and a BDD/BDD cell by electrochemical oxidation and electro-Fenton process. Journal of Electroanalytical Chemistry 728, 1-9.
Gautam, R.K., Verma, A., 2019. Electrocatalyst Materials for Oxygen Reduction Reaction in Microbial Fuel Cell. 451-483.
Gong, K., Du, F., Xia, Z., Durstock, M., Dai, L., 2009. Nitrogen-Doped Carbon Nanotube Arrays with High Electrocatalytic Activity for Oxygen Reduction. Science 323, 760-764.
Gozzi, F., Sirés, I., Thiam, A., de Oliveira, S.C., Junior, A.M., Brillas, E., 2017. Treatment of single and mixed pesticide formulations by solar photoelectro-Fenton using a flow plant. Chemical Engineering Journal 310, 503-513.
Guelfi, D.R.V., Brillas, E., Gozzi, F., Machulek, A., Jr., de Oliveira, S.C., Sires, I., 2019. Influence of electrolysis conditions on the treatment of herbicide bentazon using artificial UVA radiation and sunlight. Identification of oxidation products. Journal of environmental management 231, 213-221.
Guelfi, D.R.V., Gozzi, F., Machulek Jr, A., Sirés, I., Brillas, E., de Oliveira, S.C., 2018. Degradation of herbicide S-metolachlor by electrochemical AOPs using a boron-doped diamond anode. Catalysis Today 313, 182-188.
Guo, Y., Tong, X., Yang, N., 2023. Photocatalytic and Electrocatalytic Generation of Hydrogen Peroxide: Principles, Catalyst Design and Performance. Nano-micro letters 15, 77.
He, Q., Li, J., Qiao, Y., Zhan, S., Zhou, F., 2023. Investigation of two-electron ORR pathway of non-metallic carbon-based catalysts with P-C bond structure in Cl--bearing electrolytes. Applied Catalysis B: Environmental 339, 123087.
https://next-gen.materialsproject.org/catalysis.

https://opencatalystproject.org/.
https://www.catalysis-hub.org/.
Hu, C., Dai, L., 2019. Doping of Carbon Materials for Metal-Free Electrocatalysis. Advanced Materials 31, 1804672.
Hu, H., Gao, G.-H., Xiao, B.-B., Zhang, P., Mi, J.-L., 2023a. The oxygen reduction reaction activity and selectivity of porous-carbon supported transition metals (M-C: M Mn, Fe, Co, Ni, Cu) electrocatalysts. Diamond and Related Materials 134, 109776.
Hu, S., Zhan, Y., Wang, P., Yang, J., Wu, F., Dan, M., Liu, Z.-Q., 2023b. Urotropine-triggered multi-reactive sites in carbon nanotubes towards efficient electrochemical hydrogen peroxide synthesis. Chemical Engineering Journal 465, 142906.
Huang, S., Lu, S., Hu, Y., Cao, Y., Li, Y., Duan, F., Zhu, H., Jin, Y., Du, M., Zhang, W., Covalent Organic Frameworks with Molecularly Electronic Modulation as Metal-Free Electrocatalysts for Efficient Hydrogen Peroxide Production. Small Structures n/a, 2200387.
Iglesias, D., Giuliani, A., Melchionna, M., Marchesan, S., Criado, A., Nasi, L., Bevilacqua, M., Tavagnacco, C., Vizza, F., Prato, M., Fornasiero, P., 2018. N-Doped Graphitized Carbon Nanohorns as a Forefront Electrocatalyst in Highly Selective O2 Reduction to H2O2. Chem 4, 106-123.
Ingle, A.A., Ansari, S.Z., Shende, D.Z., Wasewar, K.L., Pandit, A.B., 2022. Progress and prospective of heterogeneous catalysts for H(2)O(2) production via anthraquinone process. Environmental science and pollution research international 29, 86468-86484.
J. K. Nørskov, J. Rossmeisl, A. Logadottir, L. Lindqvist, J. R. Kitchin, T. Bligaard, Jónsson, H., 2004. Origin of the Overpotential for Oxygen Reduction at a Fuel-Cell Cathode. J. Phys. Chem. B 104, 17886–17892.
Kamedulski, P., Skorupska, M., Koter, I., Lewandowski, M., Abdelkader-Fernandez, V.K., Lukaszewicz, J.P., 2022. Obtaining N-Enriched Mesoporous Carbon-Based by Means of Gamma Radiation. Nanomaterials 12.
Kelly, S.R., Kirk, C., Chan, K., Nørskov, J.K., 2020. Electric Field Effects in Oxygen Reduction Kinetics: Rationalizing pH Dependence at the Pt(111), Au(111), and Au(100) Electrodes. The Journal of Physical Chemistry C 124, 14581-14591.
Kim, H.W., Ross, M.B., Kornienko, N., Zhang, L., Guo, J., Yang, P., McCloskey, B.D., 2018. Efficient hydrogen peroxide generation using reduced graphene oxide-based oxygen reduction electrocatalysts. Nature Catalysis 1, 282-290.
Koh, K.H., Bagherzadeh Mostaghimi, A.H., Chang, Q., Kim, Y.J., Siahrostami, S., Han, T.H., Chen, Z., 2023. Elucidation and modulation of active sites in holey graphene electrocatalysts for H2O2 production. EcoMat 5, e12266.
Koh, K.H., Kim, Y.J., Mostaghim, A.H.B., Siahrostami, S., Han, T.H., Chen, Z., 2022. Elaborating Nitrogen and Oxygen Dopants Configurations within Graphene Electrocatalysts for Two-Electron Oxygen Reduction. ACS Materials Letters 4, 320-328.
Komtchou, S., Dirany, A., Drogui, P., Robert, D., Lafrance, P., 2017. Removal of atrazine and its by-products from water using electrochemical advanced oxidation processes. Water research 125, 91-103.
Kornienko, V.L., Kolyagin, G.A., Kornienko, G.V., Parfenov, V.A., 2020. Comparative Study of the Efficiency of New Technical Carbons CH210 and C40 in Electrosynthesis of H2O2 from O2 in Gas-Diffusion Electrodes on Their Basis. Russian Journal of Electrochemistry 56, 781-784.
Kronka, M.S., Fortunato, G.V., Mira, L., dos Santos, A.J., Lanza, M.R.V., 2023. Using Au NPs anchored on ZrO2/carbon black toward more efficient H2O2 electrogeneration in flow-by reactor for carbaryl removal in real wastewater. Chemical Engineering Journal 452, 139598.

Lai, X., Ning, X.A., Li, Y., Huang, N., Zhang, Y., Yang, C., 2023. Formation of organic chloride in the treatment of textile dyeing sludge by Fenton system. Journal of environmental sciences 125, 376-387.
Lee, J., Lee, Y., Lim, J.S., Kim, S.W., Jang, H., Seo, B., Joo, S.H., Sa, Y.J., 2023. Discriminating active sites for the electrochemical synthesis of H2O2 by molecular functionalisation of carbon nanotubes. Nanoscale 15, 195-203.
Lee, K., Lim, J., Lee, M.J., Ryu, K., Lee, H., Kim, J.Y., Ju, H., Cho, H.-S., Kim, B.-H., Hatzell, M.C., Kang, J., Lee, S.W., 2022. Structure-controlled graphene electrocatalysts for high-performance H2O2 production. Energy & Environmental Science 15, 2858-2866.
Li, C., Hu, C., Song, Y., Sun, Y.-M., Yang, W., Ma, M., 2022a. Graphene-based synthetic fabric cathodes with specific active oxygen functional groups for efficient hydrogen peroxide generation and homogeneous electro-Fenton processes. Carbon 186, 699-710.
Li, H., Kelly, S., Guevarra, D., Wang, Z., Wang, Y., Haber, J.A., Anand, M., Gunasooriya, G.T.K.K., Abraham, C.S., Vijay, S., Gregoire, J.M., Nørskov, J.K., 2021a. Analysis of the limitations in the oxygen reduction activity of transition metal oxide surfaces. Nature Catalysis 4, 463-468.
Li, M., Bai, L., Jiang, S., Sillanpää, M., Huang, Y., Liu, Y., 2023. Electrocatalytic transformation of oxygen to hydroxyl radicals via three-electron pathway using nitrogen-doped carbon nanotube-encapsulated nickel nanocatalysts for effective organic decontamination. Journal of hazardous materials 452, 131352.
Li, N., Huang, C., Wang, X., Feng, Y., An, J., 2022b. Electrosynthesis of hydrogen peroxide via two-electron oxygen reduction reaction: A critical review focus on hydrophilicity/hydrophobicity of carbonaceous electrode. Chemical Engineering Journal 450, 138246.
Li, W., Feng, Y., An, J., Yunfei, L., Zhao, Q., Liao, C., Wang, X., Liu, J., Li, N., 2022c. Thermal reduced graphene oxide enhanced in-situ H(2)O(2) generation and electrochemical advanced oxidation performance of air-breathing cathode. Environmental research 204, 112327.
Li, X., Wang, X., Xiao, G., Zhu, Y., 2021b. Identifying active sites of boron, nitrogen co-doped carbon materials for the oxygen reduction reaction to hydrogen peroxide. Journal of colloid and interface science 602, 799-809.
Li, Y., Tong, Y., Peng, F., 2020. Metal-free carbocatalysis for electrochemical oxygen reduction reaction: Activity origin and mechanism. Journal of Energy Chemistry 48, 308-321.
Ling, C., Jin, D., Li, R., Li, C., Wang, W., 2023. Self-assembled membranes modulate the active site of carbon fiber paper to boost the two-electron water oxidation reaction. Chemical Engineering Journal 465, 142903.
Liu, G., Zhou, H., Teng, J., You, S., 2019. Electrochemical degradation of perfluorooctanoic acid by macro-porous titanium suboxide anode in the presence of sulfate. Chemical Engineering Journal 371, 7-14.
Liu, J., Wei, Z., Gong, Z., Yan, M., Hu, Y., Zhao, S., Ye, G., Fei, H., 2023. Single-atom CoN4 sites with elongated bonding induced by phosphorus doping for efficient H2O2 electrosynthesis. Applied Catalysis B: Environmental 324, 122267.
Liu, W., Li, C., Ding, G., Duan, G., Jiang, Y., Lu, Y., 2022a. Highly efficient hydrogen peroxide electrosynthesis on oxidized carbon nanotubes by thermally activated-persulfate. Journal of Materiomics 8, 136-143.
Liu, Z., Gao, D., Hu, L., Liu, F., Liu, H., Li, Y., Zhang, J., Xue, Y., Tang, C., 2022b. Metal-Free Boron-Rich Borocarbonitride Catalysts for High-Efficient Oxygen Reduction to Produce Hydrogen Peroxide†. ChemistrySelect 7, e202104203.
Lu, Z., Chen, G., Siahrostami, S., Chen, Z., Liu, K., Xie, J., Liao, L., Wu, T., Lin, D., Liu, Y., Jaramillo, T.F., Nørskov, J.K., Cui, Y., 2018. High-efficiency oxygen reduction to hydrogen peroxide catalysed by oxidized carbon materials. Nature Catalysis 1, 156-162.

Lucchetti, L.E.B., Almeida, M.O., de Almeida, J.M., Autreto, P.A.S., Honorio, K.M., Santos, M.C., 2021. Density functional theory studies of oxygen reduction reaction for hydrogen peroxide generation on Graphene-Based catalysts. Journal of Electroanalytical Chemistry 895, 115429.
Luo, H., Shan, T., Zhou, J., Huang, L., Chen, L., Sa, R., Yamauchi, Y., You, J., Asakura, Y., Yuan, Z., Xiao, H., 2023. Controlled synthesis of hollow carbon ring incorporated g-C3N4 tubes for boosting photocatalytic H2O2 production. Applied Catalysis B: Environmental 337, 122933.
Ma, W., Ren, X., Li, J., Wang, S., Wei, X., Wang, N., Du, Y., Advances in Atomically Dispersed Metal and Nitrogen Co-Doped Carbon Catalysts for Advanced Oxidation Technologies and Water Remediation: From Microenvironment Modulation to Non-Radical Mechanisms. Small n/a, 2308957.
Machado, M.L.O., Paz, E.C., Pinheiro, V.S., de Souza, R.A.S., Neto, A.M.P., Gaubeur, I., dos Santos, M.C., 2022. Use of WO2.72 Nanoparticles/Vulcan® XC72 GDE Electrocatalyst Combined with the Photoelectro-Fenton Process for the Degradation of 17α-Ethinylestradiol (EE2). Electrocatalysis.
Magne, T.M., de Oliveira Vieira, T., Alencar, L.M.R., Junior, F.F.M., Gemini-Piperni, S., Carneiro, S.V., Fechine, L., Freire, R.M., Golokhvast, K., Metrangolo, P., Fechine, P.B.A., Santos-Oliveira, R., 2022. Graphene and its derivatives: understanding the main chemical and medicinal chemistry roles for biomedical applications. Journal of nanostructure in chemistry 12, 693-727.
Marques Cordeiro-Junior, P.J., Sáez Jiménez, C., Vasconcelos Lanza, M.R.d., Rodrigo Rodrigo, M.A., 2022. Electrochemical production of extremely high concentrations of hydrogen peroxide in discontinuous processes. Separation and Purification Technology 300, 121847.
Marquez, A.A., Sires, I., Brillas, E., Nava, J.L., 2020. Mineralization of Methyl Orange azo dye by processes based on H(2)O(2) electrogeneration at a 3D-like air-diffusion cathode. Chemosphere 259, 127466.
Martínez-Sánchez, C., Robles, I., Godínez, L.A., 2022. Review of recent developments in electrochemical advanced oxidation processes: application to remove dyes, pharmaceuticals, and pesticides. International Journal of Environmental Science and Technology 19, 12611-12678.
Mavrikis, S., Perry, S.C., Leung, P.K., Wang, L., Ponce de León, C., 2021. Recent Advances in Electrochemical Water Oxidation to Produce Hydrogen Peroxide: A Mechanistic Perspective. ACS Sustainable Chemistry & Engineering 9, 76-91.
Meijide, J., Gomez, J., Pazos, M., Sanroman, M.A., 2016. Degradation of thiamethoxam by the synergetic effect between anodic oxidation and Fenton reactions. Journal of hazardous materials 319, 43-50.
Miao, J., Zhu, H., Tang, Y., Chen, Y., Wan, P., 2014. Graphite felt electrochemically modified in H2SO4 solution used as a cathode to produce H2O2 for pre-oxidation of drinking water. Chemical Engineering Journal 250, 312-318.
Mohammadi, S., Zarei, M., Amini-Fazl, M.S., Ebratkhahan, M., 2023. Removal and mineralization of prednisolone from water by using homogeneous and heterogeneous electro-Fenton processes. Journal of Environmental Chemical Engineering 11, 110465.
Moraes, A., Assumpção, M.H.M.T., Simões, F.C., Antonin, V.S., Lanza, M.R.V., Hammer, P., Santos, M.C., 2015. Surface and Catalytical effects on Treated Carbon Materials for Hydrogen Peroxide Electrogeneration. Electrocatalysis 7, 60-69.
Moreira, F.C., Garcia-Segura, S., Vilar, V.J.P., Boaventura, R.A.R., Brillas, E., 2013. Decolorization and mineralization of Sunset Yellow FCF azo dye by anodic oxidation, electro-Fenton, UVA photoelectro-Fenton and solar photoelectro-Fenton processes. Applied Catalysis B: Environmental 142-143, 877-890.
Moura, J.P.C., Antonin, V.S., Trench, A.B., Santos, M.C., 2023. Hydrogen peroxide electrosynthesis: A comparative study employing Vulcan carbon modification by different MnO2 nanostructures. Electrochimica Acta 463, 142852.

Muñoz-Morales, M., Ramírez, A., Cañizares, A., Llanos, J., Ania, C., 2023. Evaluating key properties of carbon materials as cathodes for the electrogeneration of hydrogen peroxide. Carbon 210, 118082.
Murillo-Sierra, J.C., Sires, I., Brillas, E., Ruiz-Ruiz, E.J., Hernandez-Ramirez, A., 2018. Advanced oxidation of real sulfamethoxazole + trimethoprim formulations using different anodes and electrolytes. Chemosphere 192, 225-233.
Nosan, M., Strmčnik, D., Brusko, V., Kirsanova, M., Finšgar, M., Dimiev, A.M., Genorio, B., 2023. Correlating nickel functionalities to selectivity for hydrogen peroxide electrosynthesis. Sustainable Energy & Fuels 7, 2270-2278.
O. Silva, T., A. Goulart, L., Sánchez-Montes, I., O. S. Santos, G., B. Santos, R., Colombo, R., R. V. Lanza, M., 2022. Using a novel gas diffusion electrode based on PL6 carbon modified with benzophenone for efficient H2O2 electrogeneration and degradation of ciprofloxacin. Chemical Engineering Journal, 140697.
Oliveira, E.M.S., Silva, F.R., Morais, C.C.O., Oliveira, T.M.B.F., Martínez-Huitle, C.A., Motheo, A.J., Albuquerque, C.C., Castro, S.S.L., 2018. Performance of (in)active anodic materials for the electrooxidation of phenolic wastewaters from cashew-nut processing industry. Chemosphere 201, 740-748.
Ortiz-Martínez, A.K., Godínez, L.A., Martínez-Sánchez, C., García-Espinoza, J.D., Robles, I., 2021. Preparation of modified carbon paste electrodes from orange peel and used coffee ground. New materials for the treatment of dye-contaminated solutions using electro-Fenton processes. Electrochimica Acta 390, 138861.
Ortúzar, M., Esterhuizen, M., Olicón-Hernández, D.R., González-López, J., Aranda, E., 2022. Pharmaceutical Pollution in Aquatic Environments: A Concise Review of Environmental Impacts and Bioremediation Systems. Frontiers in microbiology 13, 869332.
Oturan, N., Bo, J., Trellu, C., Oturan, M.A., 2021. Comparative Performance of Ten Electrodes in Electro-Fenton Process for Removal of Organic Pollutants from Water. ChemElectroChem 8, 3294-3303.
Ouyang, R., Curtarolo, S., Ahmetcik, E., Scheffler, M., Ghiringhelli, L.M., 2018. SISSO: A compressed-sensing method for identifying the best low-dimensional descriptor in an immensity of offered candidates. Physical Review Materials 2, 083802.
Pan, G., Sun, X., Sun, Z., 2020. Fabrication of multi-walled carbon nanotubes and carbon black co-modified graphite felt cathode for amoxicillin removal by electrochemical advanced oxidation processes under mild pH condition. Environmental science and pollution research international 27, 8231-8247.
Patidar, R., Srivastava, V.C., 2022. Ultrasound Enhanced Electro-Fenton Mineralization of Benzophenone: Kinetics and Mechanistic Analysis. ACS ES&T Water 3, 1595-1609.
Paz, E.C., Aveiro, L.R., Pinheiro, V.S., Souza, F.M., Lima, V.B., Silva, F.L., Hammer, P., Lanza, M.R.V., Santos, M.C., 2018. Evaluation of H2O2 electrogeneration and decolorization of Orange II azo dye using tungsten oxide nanoparticle-modified carbon. Applied Catalysis B: Environmental 232, 436-445.
Paz, E.C., Pinheiro, V.S., Joca, J.F.S., de Souza, R.A.S., Gentil, T.C., Lanza, M.R.V., de Oliveira, H.P.M., Neto, A.M.P., Gaubeur, I., Santos, M.C., 2020. Removal of Orange II (OII) dye by simulated solar photoelectro-Fenton and stability of WO2.72/Vulcan XC72 gas diffusion electrode. Chemosphere 239, 124670.
Peleyeju, M.G., Viljoen, E.L., 2021. WO3-based catalysts for photocatalytic and photoelectrocatalytic removal of organic pollutants from water – A review. Journal of Water Process Engineering 40, 101930.

Pena-Duarte, A., Vijapur, S.H., Hall, T.D., Hayes, K.L., Larios-Rodriguez, E., Pilar-Albaladejo, J.D., Santiago, M.B., Snyder, S., Taylor, J., Cabrera, C.R., 2021. Iron Quantum Dots Electro-Assembling on Vulcan XC-72R: Hydrogen Peroxide Generation for Space Applications. ACS applied materials & interfaces 13, 29585-29601.
Peng, W., Liu, J., Liu, X., Wang, L., Yin, L., Tan, H., Hou, F., Liang, J., 2023. Facilitating two-electron oxygen reduction with pyrrolic nitrogen sites for electrochemical hydrogen peroxide production. Nature Communications 14, 4430.
Pérez-Rodríguez, S., Pastor, E., Lázaro, M.J., 2018. Electrochemical behavior of the carbon black Vulcan XC-72R: Influence of the surface chemistry. International Journal of Hydrogen Energy 43, 7911-7922.
Pérez, T., Sirés, I., Brillas, E., Nava, J.L., 2017. Solar photoelectro-Fenton flow plant modeling for the degradation of the antibiotic erythromycin in sulfate medium. Electrochimica Acta 228, 45-56.
Pinheiro, A.C.N., Bernardino, T.S., Junior, F.E.B., Lanza, M.R.V., Barros, W.R.P., 2020. Enhanced electrodegradation of the Sunset Yellow dye in acid media by heterogeneous Photoelectro-Fenton process using Fe3O4 nanoparticles as a catalyst. Journal of Environmental Chemical Engineering 8, 103621.
Pinheiro, V.S., Paz, E.C., Aveiro, L.R., Parreira, L.S., Souza, F.M., Camargo, P.H.C., Santos, M.C., 2018. Ceria high aspect ratio nanostructures supported on carbon for hydrogen peroxide electrogeneration. Electrochimica Acta 259, 865-872.
Pinheiro, V.S., Paz, E.C., Aveiro, L.R., Parreira, L.S., Souza, F.M., Camargo, P.H.C., Santos, M.C., 2019. Mineralization of paracetamol using a gas diffusion electrode modified with ceria high aspect ratio nanostructures. Electrochimica Acta 295, 39-49.
Pizzutilo, E., Kasian, O., Choi, C.H., Cherevko, S., Hutchings, G.J., Mayrhofer, K.J.J., Freakley, S.J., 2017. Electrocatalytic synthesis of hydrogen peroxide on Au-Pd nanoparticles: From fundamentals to continuous production. Chemical Physics Letters 683, 436-442.
Pointer Malpass, G.R., de Jesus Motheo, A., 2021. Recent advances on the use of active anodes in environmental electrochemistry. Current Opinion in Electrochemistry 27, 100689.
Puértolas, B., Hill, A.K., García, T., Solsona, B., Torrente-Murciano, L., 2015. In-situ synthesis of hydrogen peroxide in tandem with selective oxidation reactions: A mini-review. Catalysis Today 248, 115-127.
Pulidindi K, P.H., Hydrogen peroxide market by end-user (paper & pulp, chemical, waste water treatment, mining), industry analysis report, regional outlook, application potential, price trends, competitive market share & forecast. 2020:2020–2026.
Qian, Q., Hu, H., Huang, S., Li, Y., Lin, L., Duan, F., Zhu, H., Du, M., Lu, S., 2023. Versatile hyper-cross-linked polymer derived porous carbon nanotubes with tailored selectivity for oxygen reduction reaction. Carbon 202, 81-89.
Ramaswamy, N., Mukerjee, S., 2011. Influence of Inner- and Outer-Sphere Electron Transfer Mechanisms during Electrocatalysis of Oxygen Reduction in Alkaline Media. The Journal of Physical Chemistry C 115, 18015-18026.
Ravishankar Patil, Masirah Zahid, Sanjay Govindwar, Rahul Khandare, Govind Vyavahare, Ranjit Gurav, Neetin Desai, Soumya Pandit, Jyoti Jadhav, 2022. Chapter 8 - Constructed wetland: a promising technology for the treatment of hazardous textile dyes and effluent. Development in Wastewater Treatment Research and Processes, 173-198.
Ren, G., Lanzalaco, S., Zhou, M., Cabot, P.L., Brillas, E., Sirés, I., 2023a. Replacing carbon cloth by nickel mesh as substrate for air-diffusion cathodes: H2O2 production and carbenicillin degradation by photoelectro-Fenton. Chemical Engineering Journal 454, 140515.
Ren, P., Zhang, T., Jain, N., Ching, H.Y.V., Jaworski, A., Barcaro, G., Monti, S., Silvestre-Albero, J., Celorrio, V., Chouhan, L., Rokicińska, A., Debroye, E., Kuśtrowski, P., Van Doorslaer, S., Van Aert, S.,

Bals, S., Das, S., 2023b. An Atomically Dispersed Mn-Photocatalyst for Generating Hydrogen Peroxide from Seawater via the Water Oxidation Reaction (WOR). Journal of the American Chemical Society 145, 16584-16596.
Robles, I., Moreno-Rubio, G., García-Espinoza, J.D., Martínez-Sánchez, C., Rodríguez, A., Meas-Vong, Y., Rodríguez-Valadez, F.J., Godínez, L.A., 2020. Study of polarized activated carbon filters as simultaneous adsorbent and 3D-type electrode materials for electro-Fenton reactors. Journal of Environmental Chemical Engineering 8, 104414.
Rosa Barbosa, M.P., Lima, N.S., de Matos, D.B., Alves Felisardo, R.J., Santos, G.N., Salazar-Banda, G.R., Cavalcanti, E.B., 2018. Degradation of pesticide mixture by electro-Fenton in filter-press reactor. Journal of Water Process Engineering 25, 222-235.
Salmeron, I., Oller, I., Plakas, K.V., Malato, S., 2021. Carbon-based cathodes degradation during electro-Fenton treatment at pilot scale: Changes in H(2)O(2) electrogeneration. Chemosphere 275, 129962.
Santacruz, W., de Mello, R., Motheo, A.J., 2023. New perspectives to enhance the electro-oxidation of atrazine in methanol medium: Photo assistance using UV irradiation. Chemical Engineering Journal 466, 143093.
Santacruz, W., Fiori, I., de Mello, R., Motheo, A.J., 2022. Detection of radicals produced during electro-oxidation of atrazine using commercial DSA®-Cl2 in methanol media: Keys to understand the process. Chemosphere 307, 136157.
Santos, G., M. Cordeiro-Junior, P.J., Sánchez-Montes, I., S. Souto, R., S. Kronka, M., R.V. Lanza, M., 2022. Recent advances in H2O2 electrosynthesis based on the application of gas diffusion electrodes: Challenges and opportunities. Current Opinion in Electrochemistry 36, 101124.
Sha, P., Huang, L., Zhao, J., Wu, Z., Wang, Q., Li, L., Bu, D., Huang, S., 2023. Carbon Nitrides with Grafted Dual-Functional Ligands as Electron Acceptors and Active Sites for Ultra-stable Photocatalytic H2O2 Production. ACS Catalysis 13, 10474-10486.
Shen, W., Zhang, C., Alomar, M., Du, Z., Yang, Z., Wang, J., Xu, G., Zhang, J., Lv, J., Lu, X., 2023. Fullerene-derived boron-doped defective nanocarbon for highly selective H2O2 electrosynthesis. Nano Research.
Shi, K., Wang, Y., Xu, A., Zhou, X., Zhu, H., Wei, K., Liu, X., Shen, J., Han, W., 2021. Efficient degradation of ibuprofen by electro-Fenton with microtubular gas- diffusion electrodes synthesized by wet-spinning method. Journal of Electroanalytical Chemistry 897, 115615.
Shi, K., Wang, Y., Xu, A., Zhu, H., Gu, L., Liu, X., Shen, J., Han, W., Wei, K., 2022. Integrated electro-Fenton system based on embedded U-tube GDE for efficient degradation of ibuprofen. Chemosphere 311, 137196.
Shin, Y.-U., Yoo, H.-Y., Ahn, Y.-Y., Kim, M.S., Lee, K., Yu, S., Lee, C., Cho, K., Kim, H.-i., Lee, J., 2019. Electrochemical oxidation of organics in sulfate solutions on boron-doped diamond electrode: Multiple pathways for sulfate radical generation. Applied Catalysis B: Environmental 254, 156-165.
Siahrostami, S., Verdaguer-Casadevall, A., Karamad, M., Deiana, D., Malacrida, P., Wickman, B., Escudero-Escribano, M., Paoli, E.A., Frydendal, R., Hansen, T.W., Chorkendorff, I., Stephens, I.E., Rossmeisl, J., 2013. Enabling direct H2O2 production through rational electrocatalyst design. Nature materials 12, 1137-1143.
Sires, I., Brillas, E., 2012. Remediation of water pollution caused by pharmaceutical residues based on electrochemical separation and degradation technologies: a review. Environment international 40, 212-229.
Sires, I., Brillas, E., Oturan, M.A., Rodrigo, M.A., Panizza, M., 2014. Electrochemical advanced oxidation processes: today and tomorrow. A review. Environmental science and pollution research international 21, 8336-8367.

Skoumal, M., Rodríguez, R.M., Cabot, P.L., Centellas, F., Garrido, J.A., Arias, C., Brillas, E., 2009. Electro-Fenton, UVA photoelectro-Fenton and solar photoelectro-Fenton degradation of the drug ibuprofen in acid aqueous medium using platinum and boron-doped diamond anodes. Electrochimica Acta 54, 2077-2085.
Soares, B.S., de Mello, R., Motheo, A.J., 2023. Groundwater treatment and disinfection by electrochemical advanced oxidation process: Influence of the supporting electrolyte and the nature of the contaminant. Applied Research, 1-11.
Soto, P.C., Salamanca-Neto, C.A.R., Moraes, J.T., Sartori, E.R., Bessegato, G.G., Lopes, F., Almeida, L.C., 2020. A novel sensing platform based on self-doped TiO2 nanotubes for methylene blue dye electrochemical monitoring during its electro-Fenton degradation. Journal of Solid State Electrochemistry 24, 1951-1959.
Stacy, J., Regmi, Y.N., Leonard, B., Fan, M., 2017. The recent progress and future of oxygen reduction reaction catalysis: A review. Renewable and Sustainable Energy Reviews 69, 401-414.
Sun, L., Sun, L., Huo, L., Zhao, H., 2023a. Promotion of the Efficient Electrocatalytic Production of H2O2 by N,O- Co-Doped Porous Carbon. Nanomaterials 13, 1188.
Sun, L., Sun, L., Huo, L., Zhao, H., 2023b. Promotion of the Efficient Electrocatalytic Production of H(2)O(2) by N,O- Co-Doped Porous Carbon. Nanomaterials 13.
Syafrudin, M., Kristanti, R.A., Yuniarto, A., Hadibarata, T., Rhee, J., Al-onazi, W.A., Algarni, T.S., Almarri, A.H., Al-Mohaimeed, A.M., 2021. Pesticides in Drinking Water—A Review. International Journal of Environmental Research and Public Health 18, 468.
Tamtaji, M., Gao, H., Hossain, M.D., Galligan, P.R., Wong, H., Liu, Z., Liu, H., Cai, Y., Goddard, W.A., Luo, Z., 2022. Machine learning for design principles for single atom catalysts towards electrochemical reactions. Journal of Materials Chemistry A 10, 15309-15331.
Tao, L., Yang, Y., Yu, F., 2020. Highly efficient electro-generation of H2O2by a nitrogen porous carbon modified carbonaceous cathode during the oxygen reduction reaction. New Journal of Chemistry 44, 15942-15950.
Teng, Z., Cai, W., Sim, W., Zhang, Q., Wang, C., Su, C., Ohno, T., 2021. Photoexcited single metal atom catalysts for heterogeneous photocatalytic H2O2 production: Pragmatic guidelines for predicting charge separation. Applied Catalysis B: Environmental 282, 119589.
Thiam, A., Sires, I., Centellas, F., Cabot, P.L., Brillas, E., 2015. Decolorization and mineralization of Allura Red AC azo dye by solar photoelectro-Fenton: Identification of intermediates. Chemosphere 136, 1-8.
Trench, A.B., Moura, J.P.C., Antonin, V.S., Gentil, T.C., Lanza, M.R.V., Santos, M.C., 2023. Using a novel gas diffusion electrode based on Vulcan XC-72 carbon modified with Nb2O5 nanorods for enhancing H2O2 electrogeneration. Journal of Electroanalytical Chemistry 946, 117732.
Trevelin, L.C., Valim, R.B., Carneiro, J.F., De Siervo, A., Rocha, R.S., Lanza, M.R.V., 2020. Using black carbon modified with NbMo and NbPd oxide nanoparticles for the improvement of H2O2 electrosynthesis. Journal of Electroanalytical Chemistry 877, 114746.
Trevelin, L.C., Valim, R.B., Lourenço, J.C., De Siervo, A., Rocha, R.S., Lanza, M.R.V., 2023. Using ZrNb and ZrMo oxide nanoparticles as catalytic activity boosters supported on Printex L6 carbon for H2O2 production. Advanced Powder Technology 34, 104108.
Tu, S., Ning, Z., Duan, X., Zhao, X., Chang, L., 2022. Efficient electrochemical hydrogen peroxide generation using TiO2/rGO catalyst and its application in electro-Fenton degradation of methyl orange. Colloids and Surfaces A: Physicochemical and Engineering Aspects 651, 129657.
Valim, R.B., Trevelin, L.C., Sperandio, D.C., Carneiro, J.F., Santos, M.C., Rodrigues, L.A., Rocha, R.S., Lanza, M.R.V., 2021. Using carbon black modified with Nb2O5 and RuO2 for enhancing selectivity toward H2O2 electrogeneration. Journal of Environmental Chemical Engineering 9, 106787.

van Dommele, S., de Jong, K.P., Bitter, J.H., 2006. Nitrogen-containing carbon nanotubes as solid base catalysts. Chemical communications, 4859-4861.
Vigil-Castillo, H.H., Ruiz-Ruiz, E.J., Lopez-Velazquez, K., Hinojosa-Reyes, L., Gaspar-Ramirez, O., Guzman-Mar, J.L., 2023. Assessment of photo electro-Fenton and solar photo electro-Fenton processes for the efficient degradation of asulam herbicide. Chemosphere 338, 139585.
Wan, J., Zhang, G., Jin, H., Wu, J., Zhang, N., Yao, B., Liu, K., Liu, M., Liu, T., Huang, L., 2022. Microwave-assisted synthesis of well-defined nitrogen doping configuration with high centrality in carbon to identify the active sites for electrochemical hydrogen peroxide production. Carbon 191, 340-349.
Wang, C., Niu, J., Yin, L., Huang, J., Hou, L.-A., 2018. Electrochemical degradation of fluoxetine on nanotube array intercalated anode with enhanced electronic transport and hydroxyl radical production. Chemical Engineering Journal 346, 662-671.
Wang, F., Zhou, Y., Lin, S., Yang, L., Hu, Z., Xie, D., 2020a. Axial ligand effect on the stability of Fe–N–C electrocatalysts for acidic oxygen reduction reaction. Nano Energy 78, 105128.
Wang, J., Chen, R., Zhang, T., Wan, J., Cheng, X., Zhao, J., Wang, X., 2020b. Technological Optimization for H2O2 Electrosynthesis and Economic Evaluation on Electro-Fenton for Treating Refractory Organic Wastewater. Industrial & Engineering Chemistry Research 59, 10364-10372.
Wang, J., Li, C., Rauf, M., Luo, H., Sun, X., Jiang, Y., 2021a. Gas diffusion electrodes for H2O2 production and their applications for electrochemical degradation of organic pollutants in water: A review. The Science of the total environment 759, 143459.
Wang, J., Wang, S., 2020. Reactive species in advanced oxidation processes: Formation, identification and reaction mechanism. Chemical Engineering Journal 401, 126158.
Wang, K., Huang, J., Chen, H., Wang, Y., Song, S., 2020c. Recent advances in electrochemical 2e oxygen reduction reaction for on-site hydrogen peroxide production and beyond. Chemical communications 56, 12109-12121.
Wang, K., Huang, J., Chen, H., Wang, Y., Song, S., 2020. Recent advances on electrochemical 2e oxygen reduction reaction for hydrogen peroxide on-site production and beyond. Chemical Communications. doi:10.1039/d0cc05156j Chemical communications 56, 12109-12121.
Wang, K., Zhu, Z., Xu, D., Li, M., Yuan, S., Wang, H., 2022a. Highly active and cheap graphite/polytetrafluoroethylene composite coating cathodes for electrogeneration of hydrogen peroxide. Clean Technologies and Environmental Policy 24, 2407-2417.
Wang, N., Ma, S., Zuo, P., Duan, J., Hou, B., 2021b. Recent Progress of Electrochemical Production of Hydrogen Peroxide by Two-Electron Oxygen Reduction Reaction. Advanced science 8, e2100076.
Wang, N., Zhao, X., Zhang, R., Yu, S., Levell, Z.H., Wang, C., Ma, S., Zou, P., Han, L., Qin, J., Ma, L., Liu, Y., Xin, H.L., 2022b. Highly Selective Oxygen Reduction to Hydrogen Peroxide on a Carbon-Supported Single-Atom Pd Electrocatalyst. ACS Catalysis 12, 4156-4164.
Wang, S., Doronkin, D.E., Hahsler, M., Huang, X., Wang, D., Grunwaldt, J.D., Behrens, S., 2020d. Palladium-Based Bimetallic Nanocrystal Catalysts for the Direct Synthesis of Hydrogen Peroxide. ChemSusChem 13, 3243-3251.
Wang, W., Li, Y., Li, Y., Zhou, M., Arotiba, O.A., 2020e. Electro-Fenton and photoelectro-Fenton degradation of sulfamethazine using an active gas diffusion electrode without aeration. Chemosphere 250, 126177.
Wang, X., Zhang, Q., Jing, J., Song, G., Zhou, M., 2023a. Biomass derived S, N self-doped catalytic Janus cathode for flow-through metal-free electrochemical advanced oxidation process: Better removal efficiency and lower energy consumption under neutral conditions. Chemical Engineering Journal 466, 143283.

Wang, Y., Lin, B., 2021. Enhancement of performance for graphite felt modified with carbon nanotubes activated by KOH as Cathode in electro-fenton systems. Journal of applied biomaterials & functional materials 19, 22808000211005386.
Wang, Y., Waterhouse, G.I.N., Shang, L., Zhang, T., 2020f. Electrocatalytic Oxygen Reduction to Hydrogen Peroxide: From Homogeneous to Heterogeneous Electrocatalysis. Advanced Energy Materials 11, 2003323.
Wang, Y., Yi, M., Wang, K., Song, S., 2019. Enhanced electrocatalytic activity for H2O2 production by the oxygen reduction reaction: Rational control of the structure and composition of multi-walled carbon nanotubes. Chinese Journal of Catalysis 40, 523-533.
Wang, Y., Zhong, H., Yang, W., Feng, Y., Alonso-Vante, N., 2023b. Recent Advances with Biomass-Derived Carbon-Based Catalysts for the High-Efficiency Electrochemical Reduction of Oxygen to Hydrogen Peroxide. Advanced Energy and Sustainability Research.
Wang, Z., Zhong, W., Jiang, J., Wang, S., 2023c. Decoupling Analysis of O2 Adsorption on Metal–N–C Single-Atom Catalysts via Data-Driven Descriptors. The Journal of Physical Chemistry Letters 14, 4760-4765.
Webster, R.D., 2023. Electrochemistry combined with electron paramagnetic resonance (EPR) spectroscopy for studying catalytic and energy storage processes. Current Opinion in Electrochemistry 40, 101308.
Wei, L., Dong, Z., Chen, R., Wu, Q., Li, J., 2022a. Review of carbon-based nanocomposites as electrocatalyst for H2O2 production from oxygen. Ionics 28, 4045-4063.
Wei, Z., Deng, B., Chen, P., Zhao, T., Zhao, S., 2022b. Palladium-based single atom catalysts for high-performance electrochemical production of hydrogen peroxide. Chemical Engineering Journal 428, 131112.
Wroblowa, H., Chi-Pan, Y., Razumney, G., 1976. Electroreduction of oxygen: A new mechanistic criterion. Journal of Electroanalytical Chemistry and Interfacial Electrochemistry 69.
Wu, Z., Wang, T., Zou, J.-J., Li, Y., Zhang, C., 2022. Amorphous Nickel Oxides Supported on Carbon Nanosheets as High-Performance Catalysts for Electrochemical Synthesis of Hydrogen Peroxide. ACS Catalysis 12, 5911-5920.
Xia, G., Lu, Y., Xu, H., 2015. Electrogeneration of hydrogen peroxide for electro-Fenton via oxygen reduction using polyacrylonitrile-based carbon fiber brush cathode. Electrochimica Acta 158, 390-396.
Xia, Y., Shang, H., Zhang, Q., Zhou, Y., Hu, X., 2019. Electrogeneration of hydrogen peroxide using phosphorus-doped carbon nanotubes gas diffusion electrodes and its application in electro-Fenton. Journal of Electroanalytical Chemistry 840, 400-408.
Xu, A., Sun, X., Fan, S., Yang, Z., Zhang, Q., Zhang, Y., Zhang, Y., 2023a. Bio-FeMnOx integrated carbonaceous gas-diffusion cathode for the efficient degradation of ofloxacin by heterogeneous electro-Fenton process. Separation and Purification Technology 312, 123348.
Xu, C., Dai, L., Chen, Y., Zhang, S., He, C., Wang, X., 2023b. Enhanced interfacial interaction of mesoporous N, S co-doped carbon supported WO3-WS2 for green and selective oxidation of alcohols. Applied Surface Science 609, 155296.
Xu, H., Guo, H., Chai, C., Li, N., Lin, X., Xu, W., 2022a. Anodized graphite felt as an efficient cathode for in-situ hydrogen peroxide production and Electro-Fenton degradation of rhodamine B. Chemosphere 286, 131936.
Xu, J., Cui, Y., Wang, M., Chai, G., Guan, L., 2022b. Pyrimidine-assisted synthesis of S, N-codoped few-layered graphene for highly efficient hydrogen peroxide production in acid. Chem Catalysis 2, 1450-1466.

Xu, S., Lu, R., Sun, K., Tang, J., Cen, Y., Luo, L., Wang, Z., Tian, S., Sun, X., 2022c. Synergistic Effects in N,O-Comodified Carbon Nanotubes Boost Highly Selective Electrochemical Oxygen Reduction to H(2) O(2). Advanced science 9, e2201421.
Yan, L., Wang, C., Wang, Y., Wang, Y., Wang, Z., Zheng, L., Lu, Y., Wang, R., Chen, G., 2023. Optimizing the binding of the *OOH intermediate via axially coordinated Co-N5 motif for efficient electrocatalytic H2O2 production. Applied Catalysis B: Environmental 338, 123078.
Yang, H., Lu, N., Zhang, J., Wang, R., Tian, S., Wang, M., Wang, Z., Tao, K., Ma, F., Peng, S., 2023a. Ultra-low single-atom Pt on g-C3N4 for electrochemical hydrogen peroxide production. Carbon Energy.
Yang, S., Lu, L., Li, J., Cheng, Q., Mei, B., Li, X., Mao, J., Qiao, P., Sun, F., Ma, J., Xu, Q., Jiang, Z., 2023b. Boosting hydrogen peroxide production via establishment and reconstruction of single-metal sites in covalent organic frameworks. SusMat 3, 379-389.
Yang, S., Verdaguer-Casadevall, A., Arnarson, L., Silvioli, L., Čolić, V., Frydendal, R., Rossmeisl, J., Chorkendorff, I., Stephens, I.E.L., 2018. Toward the Decentralized Electrochemical Production of H2O2: A Focus on the Catalysis. ACS Catalysis 8, 4064-4081.
Yang, Z., Gao, Y., Zuo, L., Long, C., Yang, C., Zhang, X., 2023c. Tailoring Heteroatoms in Conjugated Microporous Polymers for Boosting Oxygen Electrochemical Reduction to Hydrogen Peroxide. ACS Catalysis 13, 4790-4798.
Ye, Z., Guelfi, D.R.V., Álvarez, G., Alcaide, F., Brillas, E., Sirés, I., 2019. Enhanced electrocatalytic production of H2O2 at Co-based air-diffusion cathodes for the photoelectro-Fenton treatment of bronopol. Applied Catalysis B: Environmental 247, 191-199.
Yu, D., Wu, M., Lin, J., 2017. Establishment of an effective activated peroxide system for low-temperature cotton bleaching using synthesized tetramido macrocyclic iron complex. Fibers and Polymers 18, 1741-1748.
Yu, F., Zhang, Y., Zhang, Y., Gao, Y., Pan, Y., 2023. Promotion of the degradation perfluorooctanoic acid by electro-Fenton under the bifunctional electrodes: Focusing active reaction region by Fe/N co-doped graphene modified cathode. Chemical Engineering Journal 457, 141320.
Yu, T., Breslin, C.B., 2020. Review—2D Graphene and Graphene-Like Materials and Their Promising Applications in the Generation of Hydrogen Peroxide. Journal of The Electrochemical Society 167, 126502.
Yue, B., Lin, L., Lei, Y., Xie, H., Si, Y., Yang, Q., Liu, X., 2023. O, N Coordination-Mediated Nickel Single-Atom Catalysts for High-Efficiency Generation of H2O2. ACS applied materials & interfaces 15, 33665-33674.
Zeng, S., Wang, S., Zhuang, H., Lu, B., Li, C., Wang, Y., Wang, G., 2022. Fluorine-doped carbon: A metal-free electrocatalyst for oxygen reduction to peroxide. Electrochimica Acta 420, 140460.
Zeng, X., Sun, J., Yao, Y., Sun, R., Xu, J.B., Wong, C.P., 2017. A Combination of Boron Nitride Nanotubes and Cellulose Nanofibers for the Preparation of a Nanocomposite with High Thermal Conductivity. ACS nano 11, 5167-5178.
Zhang, B., Xu, W., Lu, Z., Sun, J., 2020. Recent Progress on Carbonaceous Material Engineering for Electrochemical Hydrogen Peroxide Generation. Transactions of Tianjin University 26, 188-196.
Zhang, C., Yuan, L., Liu, C., Li, Z., Zou, Y., Zhang, X., Zhang, Y., Zhang, Z., Wei, G., Yu, C., 2023a. Crystal Engineering Enables Cobalt-Based Metal–Organic Frameworks as High-Performance Electrocatalysts for H2O2 Production. Journal of the American Chemical Society 145, 7791-7799.
Zhang, D., Mitchell, E., Lu, X., Chu, D., Shang, L., Zhang, T., Amal, R., Han, Z., 2023b. Metal-free carbon-based catalysts design for oxygen reduction reaction towards hydrogen peroxide: From 3D to 0D. Materials Today 63, 339-359.
Zhang, D., Zhang, H., Du, Y., Tang, H., Tang, Y., Chen, Y., Wang, Z., Sun, X., Liu, C., 2023c. Highly efficient production of hydroxyl radicals from oxygen reduction over Ni-Fe dual atom electrocatalysts for removing emerging contaminants in wastewater. Chemical Engineering Science 278, 118914.
Zhang, L., Waki, K., 2022. The influence of carboxyl group on nitrogen doping for defective carbon nanotubes toward oxygen reduction reaction. Carbon 189, 369-376.
Zhang, T., Wu, J., Wang, Z., Wei, Z., Liu, J., Gong, X., 2022a. Transfer of molecular oxygen and electrons improved by the regulation of C-N/C = O for highly efficient 2e-ORR. Chemical Engineering Journal 433, 133591.
Zhang, W., Xiao, Y., 2020. Mechanism of Electrocatalytically Active Precious Metal (Ni, Pd, Pt, and Ru) Complexes in the Graphene Basal Plane for ORR Applications in Novel Fuel Cells. Energy & Fuels 34, 2425-2434.
Zhang, X., Fu, J., Zhang, Y., Lei, L., 2008. A nitrogen functionalized carbon nanotube cathode for highly efficient electrocatalytic generation of H2O2 in Electro-Fenton system. Separation and Purification Technology 64, 116-123.
Zhang, X., Liu, J., Li, R., Jian, X., Gao, X., Lu, Z., Yue, X., 2023d. Machine learning screening of high-performance single-atom electrocatalysts for two-electron oxygen reduction reaction. Journal of colloid and interface science 645, 956-963.
Zhang, X., Zhao, X., Zhu, P., Adler, Z., Wu, Z.-Y., Liu, Y., Wang, H., 2022b. Electrochemical oxygen reduction to hydrogen peroxide at practical rates in strong acidic media. Nature Communications 13, 2880.
Zhang, Y., Wang, A., Tian, X., Wen, Z., Lv, H., Li, D., Li, J., 2016. Efficient mineralization of the antibiotic trimethoprim by solar assisted photoelectro-Fenton process driven by a photovoltaic cell. Journal of hazardous materials 318, 319-328.
Zhao, Q., An, J., Wang, S., Qiao, Y., Liao, C., Wang, C., Wang, X., Li, N., 2019. Superhydrophobic Air-Breathing Cathode for Efficient Hydrogen Peroxide Generation through Two-Electron Pathway Oxygen Reduction Reaction. ACS applied materials & interfaces 11, 35410-35419.
Zhao, Y., Hojabri, S., Sarrouf, S., Alshawabkeh, A.N., 2022. Electrogeneration of H(2)O(2) by graphite felt double coated with polytetrafluoroethylene and polydimethylsiloxane. J Environ Chem Eng 10.
Zhou, J., An, X., Lan, H., Liu, H., Qu, J., 2020. New insights into the surface-dependent activity of graphitic felts for the electro-generation of H2O2. Applied Surface Science 509, 144875.
Zhou, W., Meng, X., Gao, J., Alshawabkeh, A.N., 2019a. Hydrogen peroxide generation from O2 electroreduction for environmental remediation: A state-of-the-art review. Chemosphere 225, 588-607.
Zhou, W., Rajic, L., Meng, X., Nazari, R., Zhao, Y., Wang, Y., Gao, J., Qin, Y., Alshawabkeh, A.N., 2019b. Efficient H(2)O(2) electrogeneration at graphite felt modified via electrode polarity reversal: Utilization for organic pollutants degradation. Chem Eng J 364, 428-439.
Zhou, W., Xie, L., Gao, J., Nazari, R., Zhao, H., Meng, X., Sun, F., Zhao, G., Ma, J., 2021. Selective H2O2 electrosynthesis by O-doped and transition-metal-O-doped carbon cathodes via O2 electroreduction: A critical review. Chemical Engineering Journal 410, 128368.
Zhou, Z., Sun, M., Zhu, Y., Li, P., Zhang, Y., Wang, M., Shen, Y., 2023. A thioether-decorated triazine-based covalent organic framework towards overall H2O2 photosynthesis without sacrificial agents. Applied Catalysis B: Environmental 334, 122862.